\documentclass[twoside,twocolumn,english,british]{revtex4-1}
\usepackage[T1]{fontenc}
\usepackage[latin9]{inputenc}
\usepackage{babel}
\usepackage{array}
\usepackage{multirow}
\usepackage{amsmath}
\usepackage{amssymb}
\usepackage{graphicx}
\usepackage{wasysym}
\usepackage[unicode=true,
 bookmarks=false,
 breaklinks=true,pdfborder={0 0 0},backref=false,colorlinks=false]
 {hyperref}

\makeatletter

\providecommand{\tabularnewline}{\\}

\renewcommand{\thetable}{\arabic{table}}

\makeatother

\begin{document}

\title{DeFiNe: an optimisation-based method for robust disentangling of
filamentous networks}

\author{David Breuer$^{1,*}$ and Zoran Nikoloski$^{1}$}

\affiliation{%
\mbox{%
$^{1}$Systems Biology and Mathematical Modeling, Max Planck Institute
of Molecular Plant Physiology,%
} %
\mbox{%
Am Muehlenberg 1, 14476 Potsdam, Germany%
}}

\affiliation{$^{*}$\href{mailto:breuer@mpimp-golm.mpg.de}{breuer@mpimp-golm.mpg.de}}
\begin{abstract}
Thread-like structures are pervasive across scales, from polymeric
proteins to root systems to galaxy filaments, and their characteristics
can be readily investigated in the network formalism. Yet, network
links usually represent only parts of filaments, which, when neglected,
may lead to erroneous conclusions from network-based analyses. The
existing alternatives to detect filaments in network representations
require tuning of parameters over a large range of values and treat
all filaments equally, thus, precluding automated analysis of diverse
filamentous systems. Here, we propose a fully automated and robust
optimisation-based approach to detect filaments of consistent intensities
and angles in a given network. We test and demonstrate the accuracy
of our solution with contrived, biological, and cosmic filamentous
structures. In particular, we show that the proposed approach provides
powerful automated means to study properties of individual actin filaments
in their network context. Our solution is made publicly available
as an open-source tool, \href{http://mathbiol.mpimp-golm.mpg.de/DeFiNe/}{\textquotedblleft DeFiNe\textquotedblright},
facilitating decomposition of any given network into individual filaments. 

Keywords: polymers, cytoskeleton, networks, path cover, computational
complexity
\end{abstract}
\maketitle

\section{\label{sec:intro}Introduction}

Many network-like structures in nature are composed of filaments forming
intricate interconnected arrays across different scales of organisation.
For instance, filamentous structures can be observed in networks of
cellulose polymers in the primary cell wall of plants and algae \cite{Stamm1964,Klemm2005},
cytoskeletal networks of actin filaments or microtubules in cells
across all domains of life \cite{Shih2006,Liu2010,Wickstead2011},
networks of neurons \cite{Braitenberg1998,Lichtman2008}, root systems
\cite{Zhu2011,Galkovskyi2012,Lobet2015}, as well as solar prominences
\cite{Gibson2006,Mackay2010} and galaxy clusters \cite{Bond1996,Stoica2005,Bond2010,Tully2014}.
Network-based studies of these structures have already elucidated
important aspects such as the mechanics of cellulose networks \cite{Stamm1964,Moon2011},
transport on cytoskeletal actin networks \cite{Akkerman2011,Balint2013},
and connectivity patterns in the brain \cite{Kandel2000a,Sporns2005,Lichtman2008}.
However, the network links usually correspond to segments of the filaments;
therefore, the classical network-based analysis neglects the identities
of individual filaments. A few powerful exceptions have recently started
to emerge \cite{Xu2014,Xu2015} which may identify multiple segments
that belong to the same filament; yet, since these studies do not
capture filament overlaps, filaments are still broken into potentially
multiple fragments. Characterisation of the mechanical- \cite{Kumar2006,Bausch2006,Lu2008},
transport- \cite{Balint2013,Osunbayo2015}, and information-transmission
related properties \cite{Eccles1982,Bennett1977} in such network
representations may hence lead to erroneous conclusions due to their
differences within and between filaments. Thus, analysis of filamentous
structures rests upon accurate identification of individual filaments. 

\begin{table*}
\begin{centering}
{\small{}}%
\begin{tabular}{|l|l|>{\centering}p{2cm}|>{\centering}p{2cm}|>{\centering}p{2cm}|>{\centering}p{2cm}|>{\centering}p{2cm}|l|}
\hline 
\textbf{\small{}Input } & \textbf{\small{}Method} & \multicolumn{5}{l|}{\textbf{\small{}Features}} & \textbf{\small{}References}\tabularnewline
\hline 
 &  & \textbf{\small{}c}{\small{}urved filaments} & \textbf{\small{}f}{\small{}ilament}\\
{\small{}-specific} & \textbf{\small{}i}{\small{}ntensity}\\
{\small{}-based} & \textbf{\small{}a}{\small{}utomated} & \textbf{\small{}p}{\small{}arsi-}\\
{\small{}monious} & \tabularnewline
\hline 
\multirow{6}{*}{{\small{}image}} & {\small{}texture filter} & {\small{}$-$} & {\small{}$-$} & {\small{}$+$} & {\small{}$+$} & {\small{}$+$} & {\small{}\cite{Boudaoud2014}}\tabularnewline
\cline{2-8} 
 & {\small{}linear programming} & {\small{}$-$} & {\small{}$-$} & {\small{}$+$} & {\small{}$+$} & {\small{}$+$} & {\small{}\cite{Wood2013}}\tabularnewline
\cline{2-8} 
 & {\small{}rotating grid} & {\small{}$-$} & {\small{}$+$} & {\small{}$+$} & {\small{}$+$} & {\small{}$+$} & {\small{}\cite{Jacques2013}}\tabularnewline
\cline{2-8} 
 & {\small{}filament tracing} & {\small{}$+$} & {\small{}$+$} & {\small{}$+$} & {\small{}$\ocircle$} & {\small{}$\ocircle$} & {\small{}\cite{Cohen1994,Meijering2010,Peng2015}}\tabularnewline
\cline{2-8} 
 & {\small{}filament tracking} & {\small{}$+$} & {\small{}$+$} & {\small{}$+$} & {\small{}$+$} & {\small{}$\ocircle$} & {\small{}\cite{Mayerich2008}}\tabularnewline
\cline{2-8} 
 & open contours & {\small{}$+$} & {\small{}$+$} & {\small{}$+$} & {\small{}$+$} & {\small{}$\ocircle$} & \cite{Smith2010,Xu2015}\tabularnewline
\hline 
\multirow{2}{*}{{\small{}network}} & {\small{}rule-based decomp.} & {\small{}$+$} & {\small{}$+$} & {\small{}$-$} & {\small{}$\ocircle$} & {\small{}$\ocircle/-$} & {\small{}\cite{Leandro2009,Qiu2014}}\tabularnewline
\cline{2-8} 
 & {\small{}filament cover} & {\small{}$+$} & {\small{}$+$} & {\small{}$+$} & {\small{}$+$} & {\small{}$\ocircle/+$} & {\small{}current work}\tabularnewline
\hline 
\end{tabular}
\par\end{centering}{\small \par}

\caption{\textbf{\label{tab:intro_review}Overview of different approaches
for disentangling filamentous networks.} Two main classes of approaches
to analyse the filamentous structure of networks can be distinguished,
based on whether they operate on image data or on extracted networks.
Irrespective of the class, the existing approaches vary in their capacity
($+$) or inability ($-$) to detect curved filaments, identify individual
filaments, and to include information about the intensity/thickness
of filaments. Further, the amount of manual user input as well as
the number of parameters required by the algorithms can be feasible
($+$), laborious ($-$), or depends on the specific variant of the
algorithm ($\ocircle$). For the network-based approaches, the number
of required parameters may be different for the extraction of the
network from image data and the consequent decomposition of the network
into filaments (separated here by $/$).}
\end{table*}

Since most of the filamentous structures in natural and man-made systems
are studied by using imaging technologies, filaments are identified
either directly from the imaging data or from networks extracted from
these data (see Tab.~\ref{tab:intro_review} for succinct review).
In the first class of approaches, a texture-based method is employed
to infer the overall orientation of objects in an image section \cite{Boudaoud2014}.
However, this method cannot be employed to pinpoint individual filaments.
Another method decomposes entire images of filamentous structures
into linear segments based on a linear programming formulation \cite{Wood2013}.
While this method utilises few parameters (e.g., number of filaments),
it only models and extracts a representative set of linear filaments.
Moreover, filaments have been modelled as linear segments, detected
by co-localisation with a parallel grid at different orientations
and by using manually chosen intensity thresholds along a filament
\cite{Jacques2013}. While this method is fast and useful for extracting
linear filaments (e.g., microtubules), it does not capture bent or
tangled filaments and necessitates manual parameter selection. Alternatively,
tracing- and tracking-based methods which start from one or multiple
image points and predict neighbouring points on a putative filament
through optimisation of an energy function are powerful methods for
filament identification. Although these algorithms have led to great
insights, especially into the connectome, they typically require user
input and do not capture overlapping filaments \cite{Cohen1994,Mayerich2008,Meijering2010,Peng2015}.
Using a similar approach, open contour-based methods employ deformable
curve models that elongate and align according to an energy functional
to match the target filaments. Recent advances in open contour-based
approaches enable fully automated filament detection \cite{Xu2014,Xu2015},
but can account for the overlap of only few filaments at the expense
of parameter tuning \cite{Smith2010}. 

The second class of approaches for disentangling filamentous structures
employs a two-step procedure: First, weighted networks are extracted
from image data from different systems and imaging sources. There
is a large variety of algorithms for this task \cite{Cohen1994,Baumgarten2012,Obara2012a,Qiu2014}
which vary, in particular, in the number of parameters. Some of the
methods from the first class, presented above, may also be used to
obtain such network representations (e.g.~\cite{Mayerich2008,Xu2015}).
Second, the given, weighted networks are decomposed into filaments.
The two existing methods for this task \cite{Leandro2009,Qiu2014}
define specific junctions for bifurcations and crossings of filaments,
depending on the distances between nodes, and assign filament identities
according to manually chosen angle thresholds between incoming and
outgoing edges. In particular, they strongly restrict the potential
overlap of filaments and, due to the angle constraints, allow only
crossing but no touching filaments. Most importantly, these methods
require manual parameter selection and do not take into account filament
intensity/thickness. We note that the step of decomposing a given
network may also be beneficially applied to networks obtained, e.g.,
by open contour-based approaches in which filaments have been fragmented
due to omission of filament overlaps \cite{Xu2014,Xu2015}.

Here, we propose a robust approach to decompose a weighted network
into an optimal set of individual filaments. Therefore, our approach
addresses the second step in the second class of approaches, presented
above. The decomposition is based on a computationally difficult problem,
referred to as filament cover problem (FCP), for which we propose
suitable approximation algorithms. We test and demonstrate the accuracy
of the findings from the approximation algorithms on artificial as
well as biological and cosmic filamentous networks by comparison to
manually obtained filament covers. In addition, we demonstrate that
the proposed, fully automated solution allows facile characterisation
of well-studied properties of individual filaments, for which alternative
approaches require parameter tuning or time-consuming manual tracing.
The proposed approach is implemented in a publicly available open-source
tool, ``DeFiNe'' (\textbf{De}composing \textbf{Fi}lamentous \textbf{Ne}tworks),
which can be used to decompose any given weighted network into a set
of individual filaments for further analyses (\href{http://mathbiol.mpimp-golm.mpg.de/DeFiNe/}{http://mathbiol.mpimp-golm.mpg.de/DeFiNe/}).

\section{\label{sec:methods}Methods}

In this section we introduce the mathematical formulation of our optimisation-based
approach to decompose filamentous networks, demonstrate its computational
intractability, and formulate a suitable approximation scheme. Moreover,
we introduce new quality measures which take into account the underlying
network structures for the comparison of the obtained filament decompositions
with manual assignments used as a gold standard. Finally, we provide
a brief overview of the studied data from different biological and
physical systems. While we believe that these more technical explanations
may promote a deeper understanding of our and related approaches,
we encourage readers familiar with the aforementioned topics to proceed
directly to the Results.

\subsection{Mathematical formulation of the filament cover problem}

Any filamentous structure may be represented as weighted geometric
graph $G=\left(\mathcal{N},\mathcal{E}\right)$ with $N=\left|\mathcal{N}\right|$
nodes and $E=\left|\mathcal{E}\right|$ undirected, weighted edges.
Edges represent filament segments and their intensities or thicknesses
are reflected by their weights $w_{e}$, $e:=\left(n_{0},n_{1}\right)\in\mathcal{E}$
and $n_{0},n_{1}\in\mathcal{N}$. Nodes represent endpoints of filament
segments and their positions are denoted by $v_{n}$, $n\in\mathcal{N}$,
whereby, typically, $v_{n}\in\mathbb{R}^{2}$ or $v_{n}\in\mathbb{R}^{3}$
for networks extracted from image data. 

We naturally represent a filament by an edge-path, $p=\left(e_{p,1},\dots,e_{p,P}\right)$,
$e\in\mathcal{E}$, i.e., by an ordered sequence of $P=\left|p\right|$
adjacent edges, where $e_{p,i}$ denotes the $i$-th edge of filament
$p$. The quality of a given filament $p$ is assessed by the pairwise
filament roughness 
\begin{eqnarray}
r_{p,\mathrm{pair}} & = & \begin{cases}
\left(P-1\right)^{-1}\sum_{i=1}^{P-1}\left|w_{e_{p,i+1}}-w_{e_{p,i}}\right| & ,\,P>1\\
w_{e_{p,1}} & ,\,P=1
\end{cases},\,\,\,\,\,\,\,\,\label{eq:r_pair}
\end{eqnarray}
where $w_{e_{p,i}}$ denotes the weight of the $i$-th edge in filament
$p$. The pairwise filament roughness is small if the edge weights
along a filament vary smoothly, as expected for natural filaments
(but cf.~Discussion). For filaments that consist of one edge only,
their roughness is given by their edge weight to increase the flexibility
of our approach (cf.~Supplemental Material S1). Other roughness measures
may be readily introduced that take into account filament thicknesses
or alignments. As an additional example, we study the all-to-all filament
roughness 
\begin{eqnarray}
r_{p,\mathrm{all}} & = & \begin{cases}
\left(P-1\right)^{-1}\underset{i,j\in\left\{ 1,\dots,P\right\} }{\mathrm{max}}\left|w_{e_{p,i}}-w_{e_{p,j}}\right| & ,\,P>1\\
w_{e_{p,1}} & ,\,P=1
\end{cases},\,\,\,\,\,\,\,\,\,\label{eq:r_all}
\end{eqnarray}
which is the average maximal difference between any two edge weights
in a filament $p$, and again the original weight of the edge is used
for a filament of length one. Taking into account that most filaments
are only moderately bent, we further consider the maximal filament
deflection angle between adjacent edges of a path $p$, 
\begin{eqnarray}
r_{p,\mathrm{angle}} & = & \underset{i\in\left\{ 1,\dots,P-1\right\} }{\mathrm{max}}\label{eq:r_angle}\\
 &  & \,\,\,\mathrm{angle}\Biggl(v_{e_{p,i+1,1}}-v_{e_{p,i+1,0}},v_{e_{p,i,1}}-v_{e_{p,i,0}}\Biggr)\nonumber 
\end{eqnarray}
where $v_{e_{p,i,0}}$ and $v_{e_{p,i,1}}$ denote the positions of
the start and end nodes of the $i$-th edge of filament $p$, respectively.
Moreover, $\mathrm{angle}\left(v,v^{'}\right):=\arccos\left(\frac{v\cdot v^{'}}{\sqrt{v\cdot v}\sqrt{v^{'}\cdot v^{'}}}\right)$
is the Euclidean angle of two vectors $v$ and $v^{'}$ and $r_{p,\mathrm{angle}}=0^{\circ}$
corresponds to perfectly straight alignment. 

The optimal decomposition of a network into individual, smooth filaments
then corresponds to solving our filament cover problem (FCP; cf.~Supplemental
Material S1 for an overview of related cover problems):
\begin{quote}
Given a graph $G=\left(\mathcal{N},\mathcal{E}\right)$ and the set
$\mathcal{P}$ of all edge-paths in $G$ with roughnesses $r_{p}$,
$p\in\mathcal{P}$:

Find a subset $\mathcal{P}_{\mathrm{fil}}\subseteq\mathcal{P}$ with
minimal total (or average) roughness $R$ such that each element in
$\mathcal{E}$ is covered (at least) once. 
\end{quote}
Here, edges that are covered by more than one path naturally correspond
to filament overlaps. Minimising the average instead of the total
roughness yields shorter filaments, as appropriate for some networks
(cf.~Supplemental Material S1).

\subsection{Computational intractability of the filament cover problem and approximation
algorithm}

The FCP is computationally intractable on general and even planar
graphs (cf.~Supplemental Material S2 for motivation and proof). Graphs
generated from two-dimensional image data are planar by construction
\cite{Baumgarten2012,Obara2012a}. The proof is by reduction from
the well-studied Hamiltonian path problem which asks, for a given
network, whether there is a sequence of adjacent nodes that includes
each node exactly once, and which is known to be intractable on planar
graphs \cite{Garey1976}. Moreover, we outline an algorithm for solving
the FCP in polynomial time on trees (cf.~Supplemental Material S3). 

Since the FCP is computationally intractable on general and even planar
graphs, we devise an approximation scheme by formulating the FCP as
a fractional integer linear program (cf.~Supplemental Material S4
for motivation and details). For a given set $\mathcal{P}^{'}\subseteq\mathcal{P}$
of input paths with pairwise filament roughnesses $r_{p}$, $p\in\mathcal{P}^{'}$,
we solve:

\begin{eqnarray}
\mathrm{minimize} & \, & \frac{\sum_{p\in\mathcal{P}^{'}}r_{p}x_{p}}{\left(\sum_{p\in\mathcal{P}^{'}}x_{p}\right)^{A}}\label{eq:lp_frac}\\
\mathrm{subject\,to} & \, & \sum_{p:e\in p}x_{p}\geq1\,\mathrm{for\,all\,}e\in\mathcal{E}\nonumber \\
 & \, & x_{p}\in\left\{ 0,1\right\} \,\mathrm{for\,all\,}p\in\mathcal{P}^{'},\nonumber 
\end{eqnarray}
where we use $r_{p}\in\left\{ r_{p,\mathrm{pair}},r_{p,\mathrm{all}}\right\} $
(Eqs.~\ref{eq:r_pair} or \ref{eq:r_all}; referred to as \emph{pair
}and\emph{ all}). In the first line, $A\in\left\{ 0,1\right\} $ determines
whether the total or the average roughness is minimised (\emph{total}/\emph{avg}).
The inequality in the second line allows overlapping filaments and
equality holds for an exact cover (\emph{over}/\emph{exact}). For
$A=0$, Eq.~\ref{eq:lp_frac} is a binary linear program and for
$A=1$, the fractional problem Eq.~\ref{eq:lp_frac} may be rewritten
as a binary linear program (cf.~Supplemental Material S4). Binary
linear programs may be solved using well-established and efficient
algorithms \cite{Schrijver1998,Linderoth2005}. 

To solve the FCP for a given network, we further need to collect a
set of input paths $\mathcal{P}^{'}\subseteq\mathcal{P}$. Since for
a general graph it is not feasible to collect all paths $\mathcal{P}$
(cf.~Supplemental Material S2), we propose two approaches (referred
to as \emph{RMST} and\emph{ BFS}): (1) We create $T=100$ random minimal
spanning trees (RMST) of $G$ whose $N\left(N-1\right)/2$ non-trivial,
undirected paths are added to our set $\mathcal{P}^{'}$. To obtain
a RMST, each edge is assigned a randomly and uniformly distributed
weight and the minimum spanning tree with respect to these weights
is computed. (2) We perform a modified breadth-first search (BFS)
on the nodes, stop the search for a path $p$ when it violates the
straightness criterion $r_{p,\mathrm{angle}}<60^{\circ}$ (cf.~Eq.~\ref{eq:r_angle}),
and add all permitted paths to $\mathcal{P}^{'}$. We note that for
real-world filamentous graphs, the number of nodes and their degrees
are constrained by the filament thickness, while the number of considered
loops is further restricted by the straightness criterion, so that
our heuristically modified BFS yields a representative set $\mathcal{P}^{'}$
of paths in reasonable time. Moreover, we note that the $60^{\circ}$-criterion
is introduced for computational reasons and provides a tolerant estimate
for maximal bending of the studied real-world filaments which are
typically less bent.

\subsection{Quality assessment of filament covers via structure-aware partition
similarity measures}

The accuracy of the filaments covers obtained by solving the FCP is
assessed by comparison to manual filament assignments (cf.~Fig.~\ref{fig:res_artificial}b).
We quantify the similarity of the two partitions of the set of edges
into (potentially overlapping) filaments using the variation of information,$\mathrm{VI}$,
the Jaccard index, $\mathrm{JI}$, and the Rand index, $\mathrm{RI}$, 

\begin{eqnarray}
\mathrm{VI}\left(\mathcal{C},\mathcal{C}^{'}\right) & = & 1+\left(U\log U\right)^{-1}\cdot\label{eq:VI}\\
 &  & \,\,\,\,\,\,\,\:\cdot\sum_{i,j}g_{i,j}\left(\log\left(\frac{g_{i,j}}{g_{\cdot,j}}\right)+\log\left(\frac{g_{i,j}}{g_{i,\cdot}}\right)\right),\nonumber \\
\mathrm{RI}\left(\mathcal{C},\mathcal{C}^{'}\right) & = & \frac{h_{=,=}+h_{\neq,\neq}}{h_{=,=}+h_{=,\neq}+h_{\neq,=}+h_{\neq,\neq}},\label{eq:RI}\\
\mathrm{JI}\left(\mathcal{C},\mathcal{C}^{'}\right) & = & \frac{h_{=,=}}{h_{=,=}+h_{=,\neq}+h_{\neq,=}},\label{eq:JI}
\end{eqnarray}
where $U=\sum_{i=1}^{C}\left|\mathcal{C}_{i}\right|=\sum_{j=1}^{C^{'}}\left|\mathcal{C}_{j}^{'}\right|$,
$g_{i,j}=\left|\mathcal{C}_{i}\cap\mathcal{C}_{j}^{'}\right|$, $g_{\cdot,j}=\sum_{i=1}^{C}g_{i,j}$,
and $g_{i,\cdot}=\sum_{j=1}^{C^{'}}g_{i,j}$ \cite{Saporta2002,Meila2005,Deneud2006}.
The contingency tables $h_{\times,\times^{'}}$, $\times,\times^{'}\in\left\{ =,\neq\right\} $,
provide the numbers of edge pairs which are in the same or different
sets in the two partitions, respectively. While these classical measures
are widely used \cite{Meila2005,Lancichinetti2009}, they may generally
yield opposing results and $\mathrm{VI}$ is not well-defined for
overlapping partitions (cf.~Supplemental Material S6). More severely,
these measures do not take into account the structure of the graph
underlying the partitions. To remedy this shortcoming, we introduce
a suite of measures, the structure-aware Rand and Jaccard indices
(cf.~Eqs.~\ref{eq:RI} and \ref{eq:JI}),

\begin{eqnarray}
\mathrm{RI}^{d}\left(\mathcal{C},\mathcal{C}^{'}\right) & = & \frac{h_{=,=}^{d}+h_{\neq,\neq}^{d}}{h_{=,=}^{d}+h_{=,\neq}^{d}+h_{\neq,=}^{d}+h_{\neq,\neq}^{d}},\label{eq:RId}\\
\mathrm{JI}^{d}\left(\mathcal{C},\mathcal{C}^{'}\right) & = & \frac{h_{=,=}^{d}}{h_{=,=}^{d}+h_{=,\neq}^{d}+h_{\neq,=}^{d}}.\label{eq:JId}
\end{eqnarray}
Here $h_{\times,\times^{'}}^{d}$, $\times,\times^{'}\in\left\{ =,\neq\right\} $,
$d\in\mathbb{N}_{>0}$, count the number of edge pairs which are in
the same or different sets in the two partitions and which are separated
by at most $d$ nodes in $G$ (cf.~Supplemental Material S6 for details).
Thus, $\mathrm{RI}^{1}$ and $\mathrm{JI}^{1}$ yield structure-aware
measures of partition similarity that consider only the partition
memberships of adjacent edges (local perspective), while $\mathrm{RI}^{\infty}\equiv\mathrm{RI}$
and $\mathrm{JI}^{\infty}\equiv\mathrm{JI}$ do no not take into account
the positions of edges in the graph and reproduce the original measures
(global perspective; cf.~Supplemental Material S6 for an extensive
comparison of similarity measures and intermediates between local
and global perspective).

\subsection{Extraction of weighted networks from image data}

We test our method to disentangle filamentous networks on various
weighted, geometric networks extracted from image data. The network
extraction procedure is similar to those proposed in \cite{Baumgarten2012,Obara2012a}
(cf.~Supplemental Material S5 for details). We analyse (1) two artificial
networks extracted from drawn filamentous patterns, (2) two cytoskeletal
networks from confocal microscope images of \emph{Arabidopsis thaliana}
hypocotyl actin cytoskeletons \cite{Breuer2014}, (3) $100$ additional
cytoskeletal networks from a movie over $200\,\mathrm{s}$ from the
same experimental setup, (4) two neural networks from a fluorescence
microscopy image of a branching rat hippocampal neuron \emph{in vitro}
\cite{Brandner2014} and a schematic of a cat retinal ganglion cell
\cite{Masland2001}, respectively, and (5) two cosmic networks obtained
from images of simulated galaxy clusters \cite{Stoica2005} (see Tab.~\ref{tab:res_overview}
for an overview).

\section{\label{sec:res}Results}

\subsection{Decomposing filamentous networks is a hard optimisation problem}

A filamentous network is naturally represented as a weighted graph,
whereby the links (i.e., edges) denote segments of filaments and the
nodes represent the ends of the segments. The edge weights typically
capture the intensity or thickness of the filament segments. In this
network representation, a filament corresponds to a path given by
an ordered sequence of adjacent edges. To identify individual filaments,
we seek a decomposition of the set of edges into paths so that each
edge is covered (i.e., belongs to at least one path). Edges belonging
to more than one path naturally model filament overlaps. We will refer
to such a decomposition as a filament cover. Since a filament cover
is non-unique, we introduce a quality measure, called roughness, to
assess the quality of each path and the cover itself. Here we mainly
consider the pairwise filament roughness given by the average absolute
value of weight differences between adjacent edges. This roughness
measure quantifies how strongly the thickness varies along a filament
and is typically small for biological filaments. Disentangling the
filamentous network amounts to solving the filament cover problem
(FCP): Find a set of paths of minimum sum of roughness values that
covers the network (cf.~Methods and Supplemental Material S1 for
the mathematical formulation). The FCP formulation is quite versatile:
For instance, instead of minimising the total roughness of the filament
cover, we may minimise the average roughness. This optimisation objective
favours shorter filaments and may be more appropriate for specific
types of networks. Other roughness measures (e.g., considering the
spatial alignment of edges to penalise filaments with strong curvature)
are readily introduced and can be considered in a multi-objective
optimisation approach (cf.~Methods and Supplemental Material S1 for
different measures). 

While providing a well-defined approach towards disentangling filamentous
networks, solving the FCP is computationally prohibitive. Indeed,
we show that the FCP is intractable even on planar graphs (cf.~Methods
and Supplemental Material S2) which are used to represent filamentous
structures extracted from 2D image data \cite{Baumgarten2012,Obara2012a}.
While the FCP is solvable in polynomial time on trees (cf.~Supplemental
Material S3), most biological filamentous structures are not tree-like
as they contain loops \cite{Katifori2012,Obara2012a,Breuer2014}.
Therefore, we propose suitable approximation schemes to the FCP for
the considered networks (cf.~Methods and Supplemental Material S4
for details and the mathematical formulation). The approximation schemes
rely on collecting a large sample of paths in a given graph, followed
by the computation of the roughness of each path. The paths are collected
by performing a modified breadth-first search (BFS) or by sampling
from random minimum spanning trees (RMST). Next, we write the FCP
as classical set cover problem \cite{Karp1972} which aims at covering
the set of edges with a subset of the collected paths of minimum total
or average roughness. The set cover approximation of FCP can be formulated
and solved as a (fractional) binary linear program for which well-established
algorithms exist \cite{Schrijver1998}. The output of the program
is a set of paths which correspond to the individual filaments of
the studied network. Summarising, the FCP may be solved with different
options: The initial set of paths is obtained from a modified BFS
(denoted by \emph{BFS}) or sampling of RMSTs (\emph{RMST}), the filaments
may overlap (\emph{over}) or not (\emph{exact}), a pairwise (\emph{pair})
or all-to-all filament roughness measure (\emph{all}) is used, and
the total (\emph{total}) or average (\emph{avg}) roughness is minimised.
Since all these options are categorical, all possible $2^{4}=16$
combinations may be readily checked and no data-specific and computationally
demanding gauging of continuous parameters is necessary, as is the
case for related approaches \cite{Leandro2009,Qiu2014}. We provide
an open-source implementation of our approach, termed ``DeFiNe''
(\textbf{De}composing \textbf{Fi}lamentous \textbf{Ne}tworks), with
a simple and user-friendly graphical user interface available at \href{http://mathbiol.mpimp-golm.mpg.de/DeFiNe/}{http://mathbiol.mpimp-golm.mpg.de/DeFiNe/}.
DeFiNe takes as input a weighted graph in the standard~.gml file
format \cite{Himsolt1997} and outputs a~.gml graph with filament
identities stored as edge colours as well as a standard, human-readable~.csv-table
of various individual filament measures for custom analyses.

\begin{figure*}
\begin{centering}
\includegraphics[width=1\textwidth]{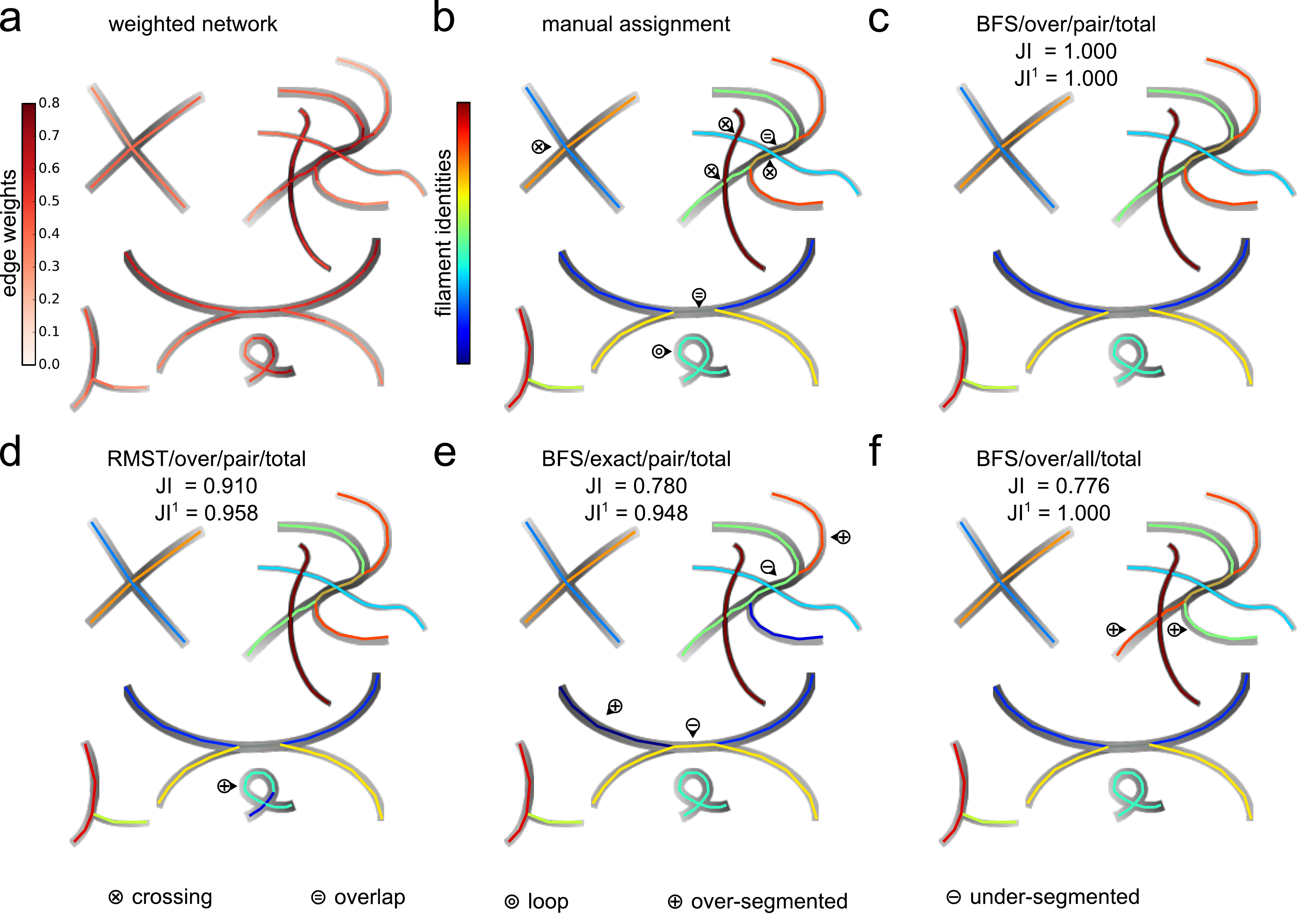}
\par\end{centering}

\caption{\textbf{\label{fig:res_artificial}Filament covers of artificial network
with crossings, overlaps, and a loop.} \textbf{(a)} Weighted, artificial
network extracted from the underlying drawing, with colour-coded edge
weights representing the local image intensity. \textbf{(b)} Manual
decomposition of the network into filaments with colour-coded indices.
The filaments display crossings ($\otimes$), overlaps ({\scriptsize{}\textcircled{=}}),
and a loop ($\circledcirc$). \textbf{(c)} Filament cover obtained
by solving the FCP using the set of input paths generated by a modified
breadth-first-search (\emph{BFS}), allowing overlapping filaments
(\emph{over}), employing the pairwise roughness measure (\emph{pair}),
and by minimising the total roughness of the cover (\emph{total}).
The automatically obtained filament cover correctly captures crossings,
overlaps, and loops, and agrees excellently with the manual assignment
(similarity of the two filament covers is measured by the global Jaccard
index, $\mathrm{JI}$, and our modified, structure-aware Jaccard index,
$\mathrm{JI}^{1}$, which reflect the fraction of pairs of all or
only adjacent edges that are assigned to the same filament, respectively;
here $\mathrm{JI}=\mathrm{JI}^{1}=1$). The filament identities and
colours are matched by solving an assignment problem whereby the total
number of edges shared by two filaments, from the manual and automated
partitioning, is maximised; the same assignment procedure is used
for the remaining panels. \textbf{(d)} When using paths obtained from
sampling random minimum spanning trees (\emph{RMST}) for the FCP,
the closed filament loop is not correctly detected and is over-segmented
($\oplus$). \textbf{(e)} When solving the exact FCP (\emph{exact}),
the loop is correctly detected. However, overlaps are neglected so
that no two filaments share an edge, leading to over- and under-segmentation
($\ominus$). \textbf{(f)} When minimising the all-to-all filament
roughness (\emph{all}), two half-filaments are interchanged because
the maximum weight difference is smaller along the altered filaments.}
\end{figure*}

\subsection{Disentangling artificial filamentous structures}

To test the accuracy of our approach, we investigate an artificial
network (Fig.~\ref{fig:res_artificial}a) of pre-specified filamentous
structure (Fig.~\ref{fig:res_artificial}b; cf.~Methods and Supplemental
Material S4 for the extraction of the network; cf. Supplemental Material
S9 for an overview of the different stages of our approach, from an
images to a network to filaments). The network contains crossing and
overlapping filaments as well as a loop (Fig.~\ref{fig:res_artificial}b,
$\otimes$, {\scriptsize{}\textcircled{=}}, and $\circledcirc$,
respectively). First, we automatically decompose the weighted filamentous
network by solving the FCP for a set of input paths from a modified
BFS, allowing for overlaps, using the pairwise roughness measure,
and minimising the total roughness of the cover (Fig.~\ref{fig:res_artificial}c,
cf.~Eq.~\ref{eq:lp_frac}). The filament identities and colours
are matched by solving an assignment problem (cf.~\cite{Kuhn1955,Wolsey1999})
such that the total number of edges shared by two filaments, from
the manual assignment and the automated cover, is maximised. The agreement
between the automated cover and the manual assignment may be measured
by classical partition similarity measures such as the Jaccard index
$\mathrm{JI}$ which counts the fraction of edge pairs which are part
of same filament \cite{Meila2005,Deneud2006}. However, $\mathrm{JI}$
does not take into account the structure of the underlying network.
Hence, we introduced a new similarity measure, $\mathrm{JI}^{1}$,
that considers only pairs of adjacent edges in each filament and thus
incorporates the network structure (cf.~Methods and and Supplemental
Material S6 for details, a generalisation to $\mathrm{JI}^{d}$ that
considers only pairs of edges which are separated by at most $d$
nodes, and a comparison of various similarity measures). For our artificial
network, solving the above FCP yields a decomposition which agrees
excellently with the manual assignment ($\mathrm{JI}=\mathrm{JI}^{1}=1$)
as all filaments are correctly detected. Second, we choose a different
set of input paths obtained from sampling RMSTs for solving th FCP
(Fig.~\ref{fig:res_artificial}d). While most filaments are correctly
detected, the loop (cf.~Fig.~\ref{fig:res_artificial}b) is over-segmented
($\oplus$) because it is not contained in the set of input paths
in its entirety (due to looplessness of trees). Third, we solve the
exact FCP which does not allow overlapping filaments (Fig.~\ref{fig:res_artificial}e).
Expectedly, the agreement with the manual assignments is lower because
filaments are over-segmented into disjoint segments and the supposedly
overlapping parts are under-segmented ($\ominus$), i.e., the respective
edges are assigned to a single filament instead of multiple filaments.
Finally, we employ the all-to-all roughness measure to assess the
quality of the filaments (Fig.~\ref{fig:res_artificial}f, cf.~Eq.~\ref{eq:r_all}).
Filament crossings, overlaps, and the loop are again correctly detected
but parts of two filaments are interchanged (cf.~$\oplus$). This
is due to the intensity/thickness of the underlying filaments which
is consistently higher for the new detected filaments which are therefore
favoured by the all-to-all roughness measure. These test cases demonstrate
the versatility and the accuracy of the proposed approach to decompose
a given network into filaments.

In the analysis of many real-world filamentous structures, the knowledge
of the underlying network structure is incomplete or the image data
impede filament detection due to low signal-to-noise ratios. To investigate
the effect of these obstacles on robust filament detection, we study
two scenarios (Supplemental Material S7): In the first scenario, we
remove a single edge from the network, recompute the optimal filament
cover, and calculate its agreement with the manual filament assignment
as measured by the structure-aware Jaccard index $\mathrm{JI}^{1}$.
We repeat the procedure for all $E$ edges and then proceed with the
removal of $E$ randomly chosen doubles of edges, triplets, up to
subsets of\foreignlanguage{english}{ }$50$ edges. As expected, the
accuracy of the filament cover typically decreases with the number
of removed edges, although removal of some specific edges even leads
to an increase in accuracy. However, $\mathrm{JI}^{1}$ decreases
very moderately by less than $0.001$ per removed edge on average
(cf.~Supplemental Material S7). In the second scenario, we assess
the robustness of our filament detection approach against noise by
adding centred Gaussian noise of increasing standard deviation to
the edge weights of the original network. For a given standard deviation,
we obtain the optimal filament covers for $100$ noisy network instances
and compute their similarity, $\mathrm{JI}^{1}$, to the manual assignment.
Again, as expected, the accuracy of the filament cover decreases with
increasing noise, but only slightly. On average, increasing the noise
by $1\%$ of the original edge weights only decreases $\mathrm{JI}^{1}$
by less than $0.001$. Moreover, we note that with increasing edge
noise the accuracy of the filament cover approaches a constant, non-zero
$\mathrm{JI}^{1}$ which reflects that some information about the
filament structure maybe obtained from the topology of the network
alone, irrespective of the edge weights (cf.~Supplemental Material
S7).

\subsection{Disentangling biological and cosmic filamentous structures}

\begin{figure*}
\begin{centering}
\includegraphics[width=1\textwidth]{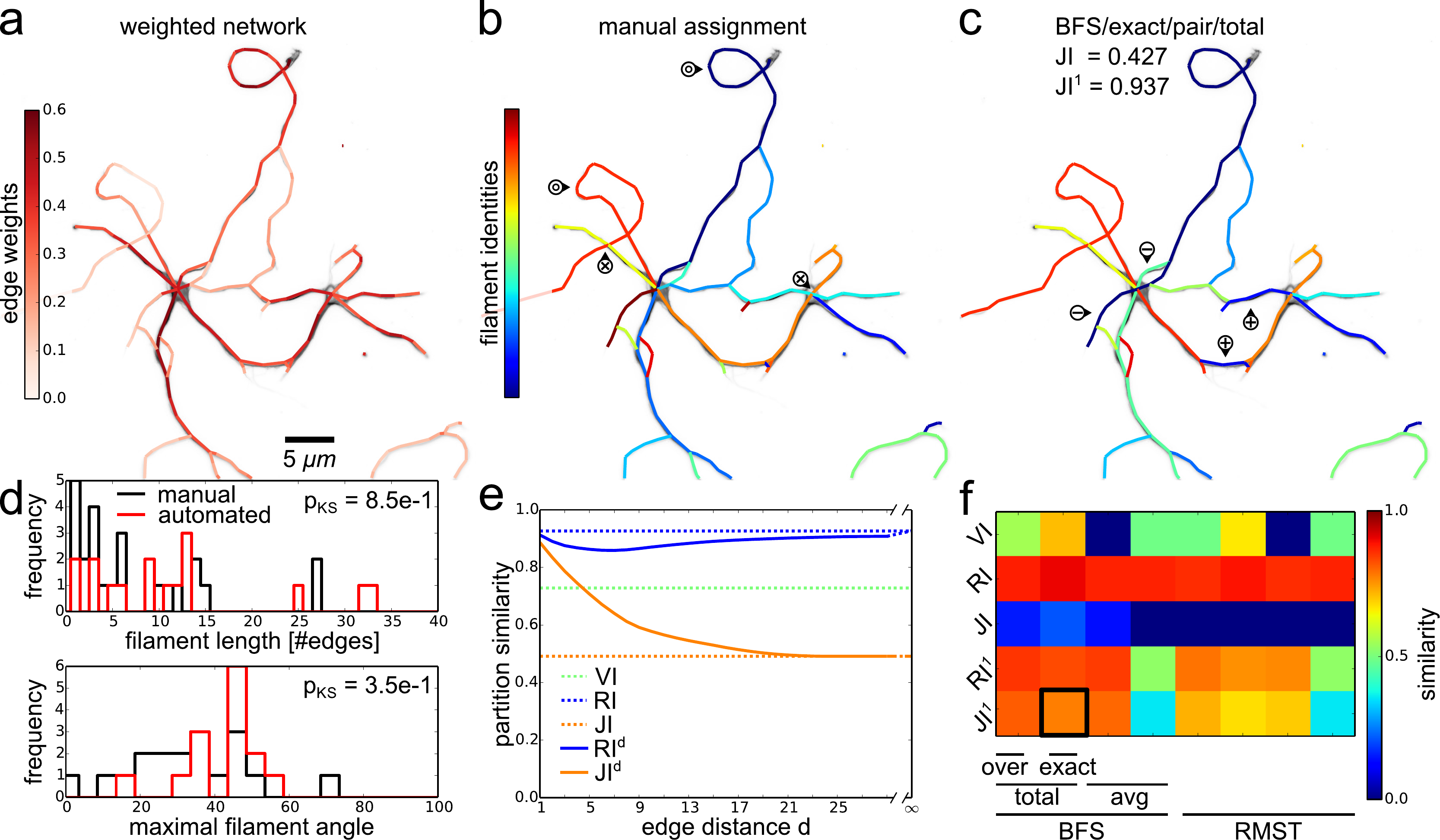}
\par\end{centering}

\caption{\label{fig:res_neuron}\textbf{Filament covers and analyses of neuronal
network.} The weighted hippocampal neuronal network is automatically
decomposed into filaments by solving the exact FCP (\emph{exact})
for paths from a modified breadth-first search (\emph{BFS}) and by
minimising the total (\emph{total}) pairwise filament roughness (\emph{pair}).
\textbf{(a)} Overlay of fluorescence microscopy image of hippocampal
neurons and extracted network with colour-coded edge weights. \textbf{(b)}
Manual decomposition of the neuronal network into filaments with colour-coded
indices and crossings ($\otimes$) and loops ($\circledcirc$). \textbf{(c)}
Automated partitioning of the network obtained by solving the FCP
displays good agreement with the manually obtained partitioning ($\mathrm{JI}^{1}$
close to $1$, see panel (e) for details) with marked illustrative
sites of over- ($\oplus$) and under-segmentation ($\ominus$). \textbf{(d)}
Distributions of numbers of edges per filament (upper panel) as well
as distributions of maximum filament angles (lower panel) are similar
for manual (black) and automated decomposition (red; Kolmogorov-Smirnov
test $p_{\mathrm{KS}}\geq0.05$). \textbf{(e)} Different measures
of similarity of manual and automated decompositions. The variation
of information $\mathrm{VI}$ (dashed green) indicates moderate similarity
but is not well-defined for general, overlapping decompositions. While
the classical Jaccard index $\mathrm{JI}$ (dashed yellow) is of small
value, the proposed structure-aware extension $\mathrm{JI}^{d}$ increases
with decreasing $d$, i.e., when only edges are considered that are
separated by at most $d$ nodes (solid yellow). Moreover, while the
classical Rand index $\mathrm{RI}$ (dashed blue) is of large value,
the proposed structure-aware extension $\mathrm{RI}^{d}$ displays
a non-monotonic dependence on $d$ (solid blue). \textbf{(f)} Heat
map of partition similarities for different similarity measures and
options of the FCP, cf.~Fig.~\ref{fig:res_artificial} for a demonstration
of the different options. The FCP options which yield the partition
shown in (c) are marked by a black rectangle.}
\end{figure*}

Since we demonstrated the power of the FCP-based approach on contrived
filamentous structures, we next proceed with investigating real biological
and cosmic filamentous structures (cf.~Methods and Supplemental Material
S5 for the extraction of the networks; cf.~Supplemental Material
S9 for an overview of the different stages of our approach). As a
first illustrative example of a biological filamentous structure,
we extract a weighted network from an image of a hippocampal neuron
(Fig.~\ref{fig:res_neuron}a) and manually obtain a filament assignment
with several crossings and loops (Fig.~\ref{fig:res_neuron}b, $\otimes$
and $\circledcirc$, respectively). Solving the FCP (same options
as in Fig.~\ref{fig:res_artificial}e) yields an automated decomposition
which captures well the manual assignment, in particular the two loops
(Fig.~\ref{fig:res_neuron}c, $\mathrm{JI}^{1}=0.937$). This is
further supported by the distributions of filament lengths (measured
by the numbers of edges) as well as the distributions of maximal filament
angles (measured between adjacent edges), which are statistically
indistinguishable between the manual assignment and the automated
decomposition (Fig.~\ref{fig:res_neuron}d, black and red; Kolmogorov-Smirnov
test $p$-value $p_{\mathrm{KS}}\geq0.05$). A detailed analysis of
the similarity of manual and automated decompositions shows that the
classical Rand index $\mathrm{RI}$ \cite{Hubert1985} overestimates
the similarity, while the variation of information $\mathrm{VI}$
\cite{Meila2003} and the Jaccard index $\mathrm{JI}$ severely underestimate
the similarity between the manual and automated decomposition when
compared to the values of the here-proposed $\mathrm{RI}^{1}$ and
$\mathrm{JI}^{1}$(Fig.~\ref{fig:res_neuron}e, dotted blue, green,
and yellow). The latter measures take into consideration the network
structure when comparing two network decompositions (Fig.~\ref{fig:res_neuron}e,
solid blue and yellow). We would like to emphasise that the disparities
in the estimations of $\mathrm{RI}$ and $\mathrm{JI}$ result from
the consideration of distant, non-adjacent edges which are excluded
in $\mathrm{RI}^{1}$ and $\mathrm{JI}^{1}$. In addition, we observe
that $\mathrm{RI}^{d}$ and $\mathrm{JI}^{d}$ show a non-trivial
dependence on the distance, $d$, between the considered edges, and
coincide with the classical similarity measure for large enough distances,
i.e., $\mathrm{RI}^{\infty}\equiv\mathrm{RI}$ and $\mathrm{JI}^{\infty}\equiv\mathrm{JI}$
(cf.~Supplemental Material S6 for a detailed discussion). 

Finally, different flavours of the FCP may be solved , as mentioned
above, to obtain decompositions of varying similarity in comparison
to the manual assignment (Fig.~\ref{fig:res_neuron}f). Solving the
FCP with paths from the modified BFS, instead of RMSTs, yields consistently
higher $\mathrm{RI}^{1}$- and $\mathrm{JI}^{1}$-values for the agreement
with the manual assignment. This is due to the higher flexibility
with respect to the treatment of loops. For the studied networks,
a decomposition based on the minimisation of the total roughness yields
higher $\mathrm{RI}^{1}$- and $\mathrm{JI}^{1}$-values in comparison
to the minimisation of the average roughness. In addition, in terms
of $\mathrm{RI}^{1}$ and $\mathrm{JI}^{1}$, covers allowing for
overlaps yield better agreement with the manual assignment, in comparison
to those in which each edge is covered by a single path. However,
these expected trends are absent or even reversed for the classical
similarity measures $\mathrm{VI}$, $\mathrm{RI}$, and $\mathrm{JI}$
(cf.~Supplemental Material S6), which further justifies the usage
of the here-proposed $\mathrm{RI}^{1}$ and $\mathrm{JI}^{1}$ for
comparing decompositions of networks arising in other network-based
analyses (cf.~e.g.~\cite{Newman2012}).

\begin{figure*}
\begin{centering}
\includegraphics[width=1\textwidth]{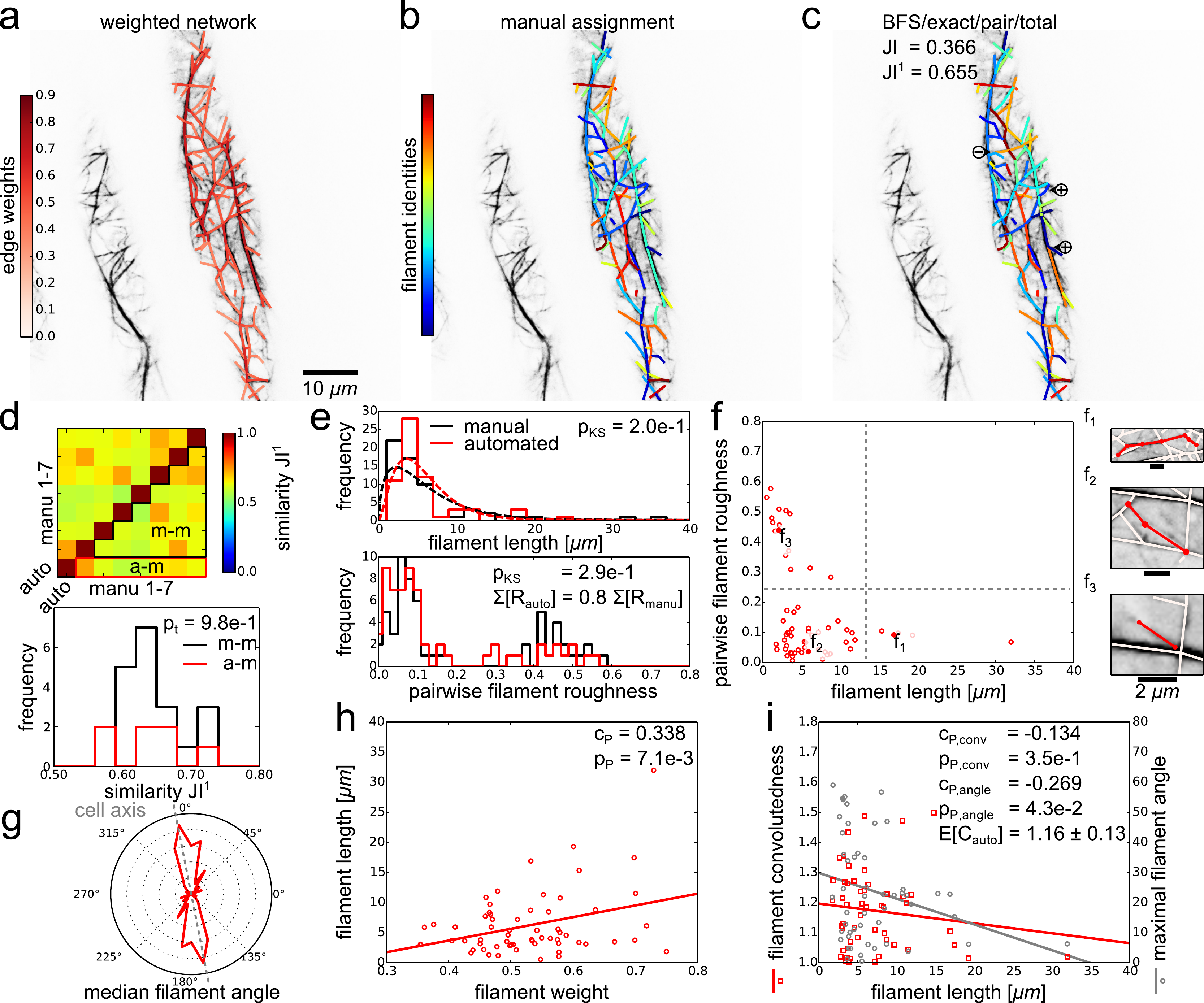}
\par\end{centering}

\caption{\textbf{\label{fig:res_actin}Filament covers and analyses of cytoskeletal
network.} The weighted cytoskeletal network is decomposed automatically
by solving the exact FCP (\emph{exact}) for paths from a modified
breadth-first search (\emph{BFS}) and by minimising the total (\emph{total})
pairwise filament roughness (\emph{pair}).\textbf{ (a)} Overlay of
confocal microscopy image of an actin cytoskeleton and extracted network
with colour-coded edge weights. \textbf{(b)} Manual decomposition
of the actin cytoskeleton into filaments with colour-coded indices.
\textbf{(c)} The automated decomposition according to the FCP correctly
assigns many of the filaments ($\mathrm{JI}^{1}=0.655$). Some occurrences
of over- ($\oplus$) and under-segmentation ($\ominus$) are marked.
\textbf{(d) }Heat map of similarity between automated (cf.~(c)) and
seven manual decompositions (cf.~e.g.~(b); upper panel). The similarities
between automated and manual decompositions (red, denoted by a-m)
do not differ from similarities among the different manual decompositions
(black, m-m; lower panel; cf.~independent two-sample Student's $t$-test
$p$-value $p_{t}\geq0.05$). \textbf{(e)} Distribution of filament
lengths for the manual (black) and automated solution (red) are similar
(upper panel; cf.~Kolmogorov-Smirnov test $p$-value $p_{\mathrm{KS}}\geq0.05$).
Maximum likelihood fits of gamma functions are shown as dashed lines.
The distributions of pairwise filament roughnesses are similar (lower
panel; cf.~$p_{\mathrm{KS}}\geq0.05$), while the total roughness
is smaller (cf.~summed $R$-values) for the automated decomposition
since it is minimised by the FCP. \textbf{(f) }Scatter plot of pairwise
filament roughness versus filament length displays three regions,
with representative examples $\mathrm{f_{1}}-\mathrm{f_{3}}$ (solid
dots): ($\mathrm{f_{1}}$) For long filaments ($\geq15\,\mathrm{\mu m}$),
the roughness is moderate ($<0.2$), as expected for actin bundles;
($\mathrm{f_{2}}$) The majority of filaments is short ($<15\,\mathrm{\mu m}$)
and of moderate roughness; ($\mathrm{f_{3}}$) Some typically short
filaments show a high roughness ($\geq0.2$), namely those which are
composed of one network edge only so that their roughness is given
by the edge weight itself (cf.~Eq.~\ref{eq:r_pair}). \textbf{(g)}
The distribution of median filament angles shows that the majority
of filaments is aligned parallel to the cell axis (grey dashed line).
\textbf{(h)} The filament length correlates with the filament weight
(cf.~linear regression and Pearson correlation coefficient $c_{P}>0$
and $p$-value $p_{\mathrm{P}}<0.05$) \textbf{(i)} Scatter plot of
filament convolutedness versus filament length shows a negative but
non-significant correlation (cf.~red squares, $c_{P,\mathrm{conv}}<0$,
and $p_{\mathrm{P},\mathrm{conv}}\geq0.05$) with an average convolutedness
of $\mathrm{E}\left[C\right]=1.16\pm0.13$. The maximum filament angle
correlates negatively and significantly with the filament length (cf.~grey
circles, $c_{P,\mathrm{angle}}<0$, and $p_{\mathrm{P},\mathrm{angle}}<0.05$),
indicating that longer (and thicker, cf.~(g)) filaments are less
curved.}
\end{figure*}

As a second biological example, we investigate the filamentous structure
of a plant actin cytoskeleton (Fig.~\ref{fig:res_actin}a). We create
seven manual assignments (one of which is shown in Fig.~\ref{fig:res_actin}b)
for a quantitative comparison with the automated decomposition (Fig.~\ref{fig:res_actin}c,
$\mathrm{JI}^{1}=0.655$; same options of the FCP as in Fig.~\ref{fig:res_artificial}e).
The agreement of the automated decomposition with the manual assignment
is good, despite several over- or under-segmented filaments (Fig.~\ref{fig:res_actin}c,
cf.~$\oplus$ and $\ominus$). For a comprehensive assessment of
this agreement, we compute the pairwise similarities between the automated
and all seven manual filament decompositions (Fig.~\ref{fig:res_actin}d,
upper panel). By comparing the similarities between automated and
manual decompositions to the similarities among the different manual
decompositions, we find reassuringly that our automated solution is
as good as any manual decomposition (Fig.~\ref{fig:res_actin}d,
lower panel, red and black, respectively; cf.~independent two-sample
Student's $t$-test $p$-value $p_{t}\geq0.05$). The agreement between
the automated decomposition and the reference manual assignment (cf.~Fig.~\ref{fig:res_actin}b)
is further confirmed by statistical tests which demonstrate that the
two distributions of filament lengths from manual assignment and automated
decomposition do not statistically differ (Fig.~\ref{fig:res_actin}e,
upper panel, black and red histograms; cf.~$p_{\mathrm{KS}}\geq0.05$).
In addition, our results indicate that the filament lengths may be
described by a gamma distribution (Fig.~\ref{fig:res_actin}e, upper
panel, dashed lines; maximal likelihood fits of normal, Weibull, and
Rayleigh distributions yield higher values for the Akaike information
criterion \cite{Akaike1974}), in agreement with theoretical and experimental
studies \cite{Burlacu1992,Ermentrout1998}. Moreover, the distributions
of average pairwise filament roughnesses do not differ between manual
assignment and automated decomposition (Fig.~\ref{fig:res_actin}e,
lower panel; cf.~$p_{\mathrm{KS}}\geq0.05$). We note that the sum
of filament roughnesses, $R$, is larger for the manual assignment
of filaments than in the automated decomposition, as expected, as
$R$ is the objective function of the minimisation in the FCP-based
formulation. 

By investigating the relationship between filament length and pairwise
roughness, we can distinguish three regions (Fig.~\ref{fig:res_actin}f):
Long filaments typically correspond to actin bundles and exhibit small
roughnesses (Fig.~\ref{fig:res_actin}$\mathrm{f_{1}}$), the majority
of filaments is shorter with comparable roughnesses (Fig.~\ref{fig:res_actin}$\mathrm{f_{2}}$),
and some typically short filaments consist of only one edge with roughness
given by the edge weight itself (Fig.~\ref{fig:res_actin}$\mathrm{f_{3}}$;
cf.~Eq.~\ref{eq:r_pair}). The angular distribution of filaments
indicates that the majority of filaments is aligned parallel to the
cell axis (Fig.~\ref{fig:res_actin}f, dashed grey line) which has
been suggested to support longitudinal cell growth \cite{Waller1997,Sampathkumar2011}.
While these reports of longitudinal alignment of the actin cytoskeleton
were based on manual or qualitative measurements, our approach facilitates
fully automated quantification of the alignment of individual filaments.
Our findings show that the length of a filament correlates with its
average weight (Fig.~\ref{fig:res_actin}g; Pearson correlation coefficient
$c_{\mathrm{P}}>0$ and $p$-value $p_{\mathrm{P}}<0.05$), i.e.,
thicker actin bundles stretch across the cell while individual thinner
actin filaments are more locally confined, as expected \cite{Staiger2009,Akkerman2011}. 

Finally, we study filament convolutedness, given by the ratio of the
length of a filament and the largest side of a bounding box enclosing
the filament, used as a measure for the curvedness of a filament \cite{Staiger2009}.
We find that the convolutedness is slightly negatively correlated
with the filament length (Fig.~\ref{fig:res_actin}i, red; $c_{\mathrm{P},\mathrm{conv}}<0$
and $p_{\mathrm{P},\mathrm{conv}}\geq0.05$), in agreement with previous
findings in \emph{Arabidopsis} \emph{thaliana} pollen grain \cite{Staiger2009}
and other plant species \cite{Henty-Ridilla2013}. In contrast to
the automated approach used here, the existing studies of filament
convolutedness required manual segmentation which may be biased by
the user. Generally, and more severely, using a bounding rectangle
to compute the convolutedness of a filament is biased by the orientation
of the filament with respect to the x- and y-axis of the image. Therefore,
we use the maximal filament angle as a non-biased measure for the
maximal, local curvedness of a filament. By investigating the relation
between the maximal filament angle and filament length, we find a
significant negative correlation (Fig.~\ref{fig:res_actin}i, grey;
$c_{\mathrm{P},\mathrm{angle}}<0$ and $p_{\mathrm{P},\mathrm{angle}}<0.05$).
This negative correlation reflects the known increase in stiffness
of actin bundles with increasing bundledness and length \cite{Gardel2004,Claessens2006}.
Thus, our approach provides a fast means to investigate this property
for individual filaments in a cellular context without laborious manual
filament identification.

To further extend these findings, we extract the cytoskeletal networks
from $100$ frames of a movie of a plant actin cytoskeleton (cf.~Methods).
For each frame, we compute the optimal filament covers and analyse
the filaments. The additional data support our reported findings (Supplemental
Material S8).

Moreover, we repeat our analyses of the robustness of our approach
against incomplete knowledge of the underlying network structure or
noisy edge weights for the cytoskeletal network (cf.~discussion of
Fig.~\ref{fig:res_artificial}; Supplemental Material S7). In our
first scenario, the removal of increasing numbers of edges typically
moderately decreases the accuracy of the obtained filament covers,
i.e., their agreement with the manual assignment as measured by $\mathrm{JI}^{1}$.
While the removal of some critical edges leads to a more severe decrease
in accuracy, there exist edges whose removal leads to an increase
in accuracy. On average, the removal of one additional edge decreases
$\mathrm{JI}^{1}$ by around $0.002$. Consequently, a loss of $10\%$
of the cytoskeletal network's $E=179$ edges still yields $\mathrm{JI}^{1}\approx0.6$
which is comparable to similarity values between different manual
assignments (cf.~Fig.~\ref{fig:res_actin}d; cf.~Supplemental Material
S7). In our second scenario, the adding of Gaussian noise of increasing
standard deviation to the edge weights similarly, as expected, decreases
the accuracy of the obtained filament covers. However, this effect
is moderate, i.e, increasing the standard deviation by $1\%$ of the
original edge weight decreases $\mathrm{JI}^{1}$ by less than $0.001$.
Adding noise with a standard deviation of $20\%$ of the original
edges weights still yields $\mathrm{JI}^{1}\approx0.6$. As for the
robustness analyses of the contrived network, for strong noise, $\mathrm{JI}^{1}$
tends to a constant, non-zero value which suggests that some information
about the filament structure may be obtained solely from the network
topology, irrespective of the edge weights (cf.~discussion of Fig.~\ref{fig:res_artificial};
cf.~Supplemental Material S7).

\begin{figure*}
\begin{centering}
\includegraphics[width=1\textwidth]{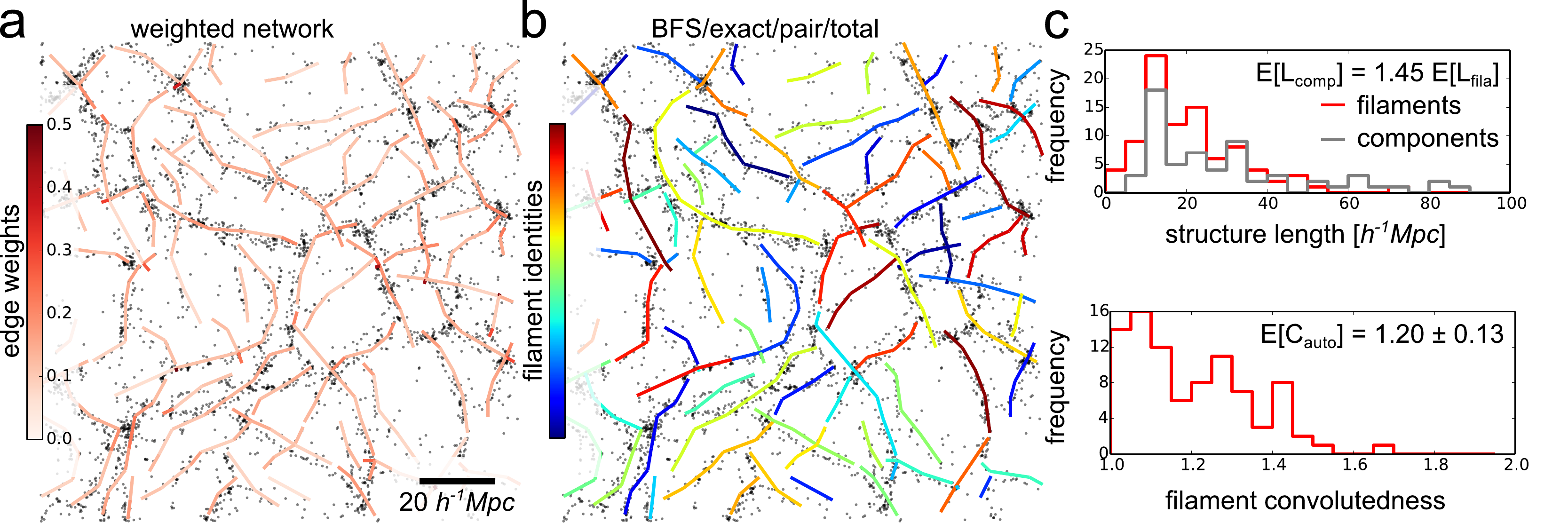}
\par\end{centering}

\caption{\textbf{\label{fig:res_galaxy}Filament covers and analyses of cosmic
web.} Image data from: Stoica et al., A\&A, 434, 423-432, 2005, reproduced
with permission {\scriptsize{}\textcircled{c}} ESO \cite{Stoica2005}.
The cosmic web is decomposed automatically by solving the exact FCP
(\emph{exact}) for paths from a modified breadth-first search (\emph{BFS})
and by minimising the total (\emph{total}) pairwise filament roughness
(\emph{pair}). Distances are given in $\mathrm{h^{-1}Mpc}$, where
currently $\mathrm{h}\approx0.7$ is the dimensionless Hubble parameter
\cite{Croton2013}.\textbf{ (a)} Overlay of simulated galaxy clusters
and extracted network with colour-coded edge weights. \textbf{(b)}
Automated decomposition of the cosmic web into galaxy filaments with
colour-coded indices. \textbf{(c) }The length distribution of galaxy
filaments exhibits a peak around $20\,\mathrm{h^{-1}Mpc}$ and levels
off for larger lengths (upper panel, red). As a comparison, the distribution
of the total lengths of the connected components levels off more slowly
and overestimates the average filament length by a factor of $1.45$
(upper panel, grey; cf.~average $L$-values). The distribution of
the convolutedness of galaxy filaments suggests a prevalence of straight
filaments and its average is comparable to that of the actin network
(cf.~\ref{fig:res_actin}i; cf.~ $\mathrm{E}\left[C\right]=1.20\pm0.13$). }
\end{figure*}

As a final example, we decompose the network of a simulated galaxy
cluster (Fig.~\ref{fig:res_galaxy}a) into individual galaxy filaments
(Fig.~\ref{fig:res_galaxy}b). The quantification of galaxy filaments
may help to elucidate the acceleration of the universe \cite{Sousbie2008a}
and improve our understanding of large-scale structure formation \cite{Sousbie2008}.
Moreover, studies have revealed gravitational motion of galaxies along
individual filaments \cite{Faltenbacher2002,Aubert2004}. Yet, previous
studies focused on connected components of the cosmic web, and sought
robust methods to identify individual filaments \cite{Stoica2005,Sousbie2008}.
Our approach confirms the expected discrepancy between the lengths
of the components (i.e., the sum of their edge lengths; Fig.~\ref{fig:res_galaxy}c,
upper panel, grey) and the length of individual filaments (Fig.~\ref{fig:res_galaxy}c,
upper panel, red; cf.~average $L$-values). Moreover, the decomposition
of the cosmic structures enables analyses of individual filament shapes.
For example, the convolutedness which measures the curvedness of a
filament shows small values (Fig.~\ref{fig:res_galaxy}, lower panel),
which are interestingly comparable to those found in the actin cytoskeleton
(cf.~Fig.~\ref{fig:res_actin}i; cf.~average $C$-values), indicating
the prevalence of straight galaxy filaments.

\begin{table*}
\begin{centering}
{\small{}}%
\begin{tabular}{|c|l|l|llll|c|c|c|c|c|}
\hline 
 &  & \textbf{\small{}Figure} & \multicolumn{4}{l|}{\textbf{\small{}Options}} & \multicolumn{5}{l|}{\textbf{\small{}Similarity}}\tabularnewline
\hline 
 &  &  & \multicolumn{4}{l|}{} & {\small{}$\mathrm{VI}$} & {\small{}$\mathrm{RI}\left(\equiv\mathrm{RI}^{\infty}\right)$} & {\small{}$\mathrm{JI}\left(\equiv\mathrm{JI}^{\infty}\right)$} & {\small{}$\mathrm{RI}^{1}$} & {\small{}$\mathrm{JI}^{1}$}\tabularnewline
\hline 
\multirow{2}{*}{{\small{}artificial}} & {\small{}overlaps + loop} & {\small{}\ref{fig:res_artificial}} & \emph{\small{}BFS} & \emph{\small{}over} & \emph{\small{}pair} & \emph{\small{}tot} & {\small{}0.792} & {\small{}1.000 } & {\small{}1.000 } & {\small{}1.000 } & {\small{}1.000 }\tabularnewline
\cline{2-12} 
 & {\small{}grid-like} & {\small{}S5b} & \emph{\small{}BFS} & \emph{\small{}exact} & \emph{\small{}pair} & \emph{\small{}tot} & {\small{}0.889 } & {\small{}0.962 } & {\small{}0.742 } & {\small{}0.941 } & {\small{}0.872 }\tabularnewline
\hline 
\multirow{2}{*}{{\small{}neural}} & {\small{}hippocampus} & {\small{}\ref{fig:res_neuron}} & \emph{\small{}BFS} & \emph{\small{}exact} & \emph{\small{}pair} & \emph{\small{}tot} & {\small{}0.848 } & {\small{}0.906 } & {\small{}0.427} & {\small{} 0.954 } & {\small{} 0.937}\tabularnewline
\cline{2-12} 
 & {\small{}retina} & {\small{}S5d} & \emph{\small{}BFS} & \emph{\small{}exact} & \emph{\small{}pair} & \emph{\small{}tot} & {\small{}0.792 } & {\small{}0.963} & {\small{}0.397} & {\small{}0.905} & {\small{}0.883}\tabularnewline
\hline 
\multirow{2}{*}{{\small{}cytoskeletal}} & {\small{}actin (FABD-labelled)} & {\small{}\ref{fig:res_actin}} & \emph{\small{}BFS} & \emph{\small{}exact} & \emph{\small{}pair} & \emph{\small{}tot} & {\small{}0.829} & {\small{} 0.976} & {\small{} 0.366} & {\small{} 0.854 } & {\small{} 0.655}\tabularnewline
\cline{2-12} 
 & {\small{}actin (Lifeact-labelled)} & {\small{}S5f} & \emph{\small{}BFS} & \emph{\small{}exact} & \emph{\small{}pair} & \emph{\small{}tot} & {\small{}0.530 } & {\small{}0.929} & {\small{} 0.193} & {\small{} 0.838 } & {\small{}0.701}\tabularnewline
\hline 
\multirow{2}{*}{{\small{}cosmic}} & {\small{}galaxy cluster (sparse)} & {\small{}\ref{fig:res_galaxy}} & \emph{\small{}BFS} & \emph{\small{}exact} & \emph{\small{}pair} & \emph{\small{}tot} & \multicolumn{5}{l|}{{\small{}~~~~~ no manual assignment}}\tabularnewline
\cline{2-7} 
 & {\small{}galaxy cluster (dense)} & {\small{}S5h} & \emph{\small{}BFS} & \emph{\small{}exact} & \emph{\small{}pair} & \emph{\small{}tot} & \multicolumn{5}{l|}{{\small{}~~~~~ for comparison}}\tabularnewline
\hline 
\end{tabular}
\par\end{centering}{\small \par}

{\small{}\caption{\textbf{\label{tab:res_overview}Quality of filament covers of artificial,
biological, and cosmic networks in comparison to manual decompositions.}
A given network is decomposed into filaments by solving the FCP with
different options: The initial set of paths is obtained from a modified
breadth-first search (\emph{BFS}) or sampling of random minimum spanning
trees (\emph{RMST}), the filaments may overlap (\emph{over}) or not
(\emph{exact}), a pairwise (\emph{pair}) or all-to-all filament roughness
measure (\emph{all}) is used, and the total (\emph{total}) or average
(\emph{avg}) roughness is minimised. The table displays the investigated
filament covers with high similarity to the manual assignments.}
}{\small \par}
\end{table*}

In Tab.~\ref{tab:res_overview}, we summarise the quality of the
investigated decompositions of different filamentous networks and
the options of the underlying FCP (cf.~Supplemental Material S8 and
S9 for analyses of additional filamentous networks that are not shown
in the main text).

\section{\label{sec:discussion}Discussion}

The decomposition of complex networks into meaningful substructures
has facilitated network-based analyses of systems found in nature
or designed by humans \cite{Milo2002,Shen-Orr2002,Sporns2004}. These
natural and technical networks often embed filaments as basic building
units. To enable deeper understanding of network systems with filamentous
structure, it is therefore paramount to develop methods for accurate
and feasible identification of the underlying filaments. In particular,
the distinction between intra- and inter-filament connections enables
a more detailed analysis of filamentous structures, including length
statistics, spatial alignment, and bending of individual filaments.
Such statistics may offer new insights, e.g., into the role of single
actin or galaxy filaments in their cellular or cosmic network context,
respectively (cf.~Figs.~\ref{fig:res_actin}e-i and \ref{fig:res_galaxy}c).

Here, we proposed a robust optimisation approach to decompose any
given weighted network into a set of smooth filaments comprising a
filament cover. Since we demonstrated that the filament cover problem
is intractable on general networks, we proposed, tested, and validated
several alternative approximation schemes. The proposed approximation
schemes are gauged at applications from different scientific fields
in which filamentous structures naturally arise. We applied our optimisation-based
approach on contrived test cases as well as biological and cosmic
networks, and showed that it reliably identifies crossing, (non-)
overlapping, and looped filaments in agreement with expert-based manual
assignments.

Our approach offers a number of advantages over the existing alternatives:
(1) The proposed optimisation approach can be applied to any weighted
network. In particular, the approach can be readily applied to any
network generated from two- or three-dimensional experimental image
data typically gathered in biological studies and analyses of man-made
systems (e.g.~\cite{Masland2001,Paredez2006,Riedl2008,Tero2010}),
irrespective of the image source (e.g., light microscopy- or MRI-based).
Thus, it may be used to study a variety of natural and technical filamentous
structures in search for universal properties which go beyond the
characterisation of geometric networks \cite{Barthelemy2011}.

(2) Our approach facilitates the establishment of a link between the
dynamics of individual filaments and the dynamics of the whole network.
While the dynamics of individual filaments is guided by typically
molecular, local processes, the behaviour of the entire filamentous
structure incorporates and responds to stimuli across different scales.
Therefore, the proposed approach provides the starting point towards
network-oriented analysis of filaments. More specifically, the filament
covers may even be used to track mobile filaments, as has been proposed
for images of a few filaments using open contours \cite{Smith2010},
providing a venue for fruitful applications of the method. 

(3) The different options of our approach, e.g., different measures
of the filament roughness, enable flexible and intuitive customisation
for different types of networks. For example, the filament roughness
measure may include a penalty for filament bending in networks of
straight, stiff filaments (such as microtubules \cite{Gittes1993,Mameren2009}),
or a penalty for length deviations in networks of filaments of mostly
uniform length (such as synthetic polymers that are used, e.g., in
drug delivery systems \cite{Ali2006,Hartmann2009}). 

(4) At the same time, our approach to disentangle a given network
is parsimonious, i.e., it has a strictly limited number of categorical
options which allow testing of all possible combinations ($4^{2}=16$
in total). In contrast, approaches which rely on multiple continuous
parameters require data-specific and computationally expensive gauging
of the parameters \cite{Leandro2009,Qiu2014}. When compared to approaches
which detect filaments directly from image data, however, the parsimony
of our approach is counterbalanced by the parameter requirements of
the preceding network extraction procedure.

(5) Nevertheless, approaches that detect filaments directly from image
data typically rely on local optimisation schemes and thus, e.g.,
on the order of filament initialisations and definitions of local
filament properties \cite{Mayerich2008,Meijering2010,Peng2015,Xu2015}.
In contrast, our approach offer the advantage that the decomposition
into filaments is performed in a single optimisation step which holistically
considers the global structure of both filaments and network.

(6) Finally, since our approach replies on a general network representation,
it may be applied also to networks obtained from other, e.g., open
contour-based methods which often do not capture filament overlaps
and result in fragmented filaments \cite{Xu2014,Xu2015}. In a post-processing
step, these fragments may be conveniently merged using our network-based
approach (cf.~Supplemental Material 10).

Yet, some caution is warranted: (1) The available options of the FCP
yield different decompositions. We showed that paths sampled from
a modified BFS enable more flexible and more accurate decompositions
in comparison to paths sampled from RMSTs (cf.~Fig.~\ref{fig:res_artificial});
in contrast to minimising the the average roughness, the minimisation
of the total roughness favours longer filaments in better accordance
with the manual assignments (cf.~Fig.~\ref{fig:res_artificial});
moreover, since filament overlaps in biological systems may lead to
an abrupt increase in apparent filament thickness, the proposed all-to-all
filament roughness may be more suitable to study such situations than
the pairwise filament roughness which favours filaments of slowly
varying thickness. Therefore, the suitable choice of feasible and
suitable options has to be further investigated. For example, for
the actin cytoskeletal networks, it is not obvious if overlapping
filaments should be preferred over non-overlapping filaments and if
the pairwise roughness is a better measure of filament quality than
the all-to-all roughness. Yet, such decision problems are innate not
only to all automated decomposition algorithms, but also to the manual
assignment based on which the performance is assessed. Thus, exploring
different decomposition options by an expert in the field may hint
at the right choice.

(2) The quality of the filament cover clearly depends on the quality
of the input network. To this end, several algorithms have been proposed
for the extraction of various types of networks from image data with
low error rates \cite{Cohen1994,Baumgarten2012,Meijering2010,Obara2012a,Qiu2014,Xu2015}.
Moreover, we investigated different scenarios to test the robustness
of our approach against incomplete knowledge of the underlying network
structure as well as low signal-to-noise ratios and found that the
accuracy of the filament cover is only moderately affected by these
obstacles (cf.~Supplemental Material S7).

(3) Another issue are the computational requirements of the FCP. Although
our proposed approximation scheme employs a modified BFS and a binary
linear program which run fast on the tested networks, it may become
infeasible for larger networks comprising more edges or nodes of larger
degrees. Therefore, future efforts may focus on devising algorithms
which approximate the FCP by employing local searches, i.e., without
sampling a large number of paths for the proposed set cover-based
approximation scheme.

(4) Finally, we note that many polymers are not simple linear chains
but branched tree-like structures \cite{Inoue2000,Tomalia2001}. Also
many neurons may be naturally described as tree-like structures \cite{Verwer1983,Ascoli2007}.
Our approach can be extended to account for these cases, thus, opening
a new field of research. To this end, covering networks with more
complex structures, such as stars \cite{Tarsi1981,Cohen1991,Lin1996}
or, more generally, trees \cite{Even2004,Horak2008} may be employed.
Due to intractability of these problems, investigation of approximation
schemes like our set cover formulation will be needed. A central question
will be the development of measures for the quality of a given star
or tree cover.

In conclusion, by decomposing technically and biologically relevant
filamentous structures into their constitutive filaments, our approach
allows to see both the wood and the trees.

\section{Acknowledgements}

D.B.~and Z.N.~acknowledge financial support by the Max Planck Society.

\section{Author contributions}

D.B.~implemented the method and analysed the data; D.B.~and Z.N.~developed
the method, showed the computational complexity of the problem, and
wrote the manuscript.

\section{Competing financial interests}

The authors declare no competing financial interests. 

\bibliographystyle{apalike}

\newpage
\newpage
\onecolumngrid
\setcounter{page}{1}
\renewcommand{\thepage}{S\arabic{page}}
\setcounter{table}{0}
\renewcommand{\thetable}{S\arabic{table}}
\setcounter{figure}{0}
\renewcommand{\thefigure}{S\arabic{figure}}
\setcounter{section}{0}
\renewcommand{\thesection}{S\arabic{section}}
\setcounter{equation}{0}
\renewcommand{\theequation}{S\arabic{equation}}

\begin{center}
\textbf{\large{}DeFiNe: an optimisation-based method for robust disentangling
of filamentous networks}
\par\end{center}{\large \par}

\begin{center}
\textbf{\large{}Supplemental Material S1-S10}
\par\end{center}{\large \par}

\begin{center}
~
\par\end{center}

\begin{center}
{\large{}David Breuer$^{1,*}$ and Zoran Nikoloski$^{1}$}
\par\end{center}{\large \par}

\begin{center}
~
\par\end{center}

\begin{center}
\emph{}%
\mbox{%
\emph{$^{1}$Systems Biology and Mathematical Modeling, Max Planck
Institute of Molecular Plant Physiology}%
}\emph{, }%
\mbox{%
\emph{Am Muehlenberg 1, 14476 Potsdam, Germany}%
}
\par\end{center}

\begin{center}
\emph{$^{*}$breuer@mpimp-golm.mpg.de}
\par\end{center}

\begin{center}
~
\par\end{center}

\section{\label{sec:app_math}Supplemental Material S1: Mathematical formulation
of the filament cover problem}

The structure of a filamentous network is described by a weighted
geometric graph $G=\left(\mathcal{N},\mathcal{E}\right)$ with $N=\left|\mathcal{N}\right|$
nodes and $E=\left|\mathcal{E}\right|$ undirected, weighted edges.
Edges represent filament segments and nodes represent their endpoints.
The positions of the nodes are $v_{n}$, $n\in\mathcal{N}$, whereby,
typically, $v_{n}\in\mathbb{R}^{2}$ or $v_{n}\in\mathbb{R}^{3}$
for networks extracted from image data. We focus on geometric networks
because filaments are embedded in space, but our approach is readily
applicable to non-geometric graphs. The edge weights are $w_{e}$,
$e:=\left(n_{0},n_{1}\right)\in\mathcal{E}$ and $n_{0},n_{1}\in\mathcal{N}$. 

To decompose the graph $G$ into individual filaments it is natural
to decompose it into paths, i.e., to solve a path cover problem (PCP).
The PCP has been intensively studied on different types of graphs
and with various restrictions (e.g.~\cite{Rao1990,Andreatta1995,Lin1995,Pak-Ken1999,Lin2006,Bresar2011}).
There are several potential routes (cf.~\cite{Andreatta1995} for
an overview of the PCP for testing printed circuits): (1) We may either
use node- or edge-paths, where a path $p=\left(a_{p,1},\dots,a_{p,P}\right)$
is an ordered sequence of $P=\left|p\right|$ pairwise adjacent nodes
($a\in\mathcal{N}$) or edges ($a\in\mathcal{E}$), respectively,
and $a_{p,i}$ denotes the $i$-th node or edge of filament $p$.
(2) The paths may be either node-disjoint, edge-disjoint, or unrestricted.
(3) The objective of the PCP may be either to obtain a cover of minimum
cardinality or minimum weight. 

For our purpose, the decomposition of a filamentous network into individual
smooth filaments, it seems reasonable to look for an edge-path cover
where each edge is covered by (at least) one path and the total (or
average) roughness is minimised. Edges that are covered by more than
one path naturally correspond to filament overlaps. The minimisation
of the average instead of the total roughness favours shorter paths
which may be appropriate for some networks. 

To define our filament cover problem (FCP) more rigorously, we introduce
the roughness $r_{p}$ of path $p$ and the set $\mathcal{P}$ of
all paths in $G$:
\begin{quote}
Given a set $\mathcal{E}$ of edges and a set $\mathcal{P}$ of paths
with roughnesses $r_{p}$, $p\in\mathcal{P}$:

Find a subset $\mathcal{P}_{\mathrm{fil}}\subseteq\mathcal{P}$ with
minimal total (or average) roughness $R$ such that each element in
$\mathcal{E}$ is covered (at least) once. 
\end{quote}
The roughness measure $r_{p}$ of a path $p$ can be chosen arbitrarily
and may involve, e.g., the edge weights or the edge alignments. An
intuitive choice is the pairwise filament roughness of $p$ (cf.~Eq.~\ref{eq:r_pair}),
\begin{eqnarray}
r_{p,\mathrm{pair}} & = & \begin{cases}
\left(P-1\right)^{-1}\sum_{i=1}^{P-1}\left|w_{e_{p,i+1}}-w_{e_{p,i}}\right| & ,\,P>1\\
w_{e_{p,1}} & ,\,P=1
\end{cases},\,\,\,\,\,\,\,\,\label{eq:supp_r_pair}
\end{eqnarray}
where $w_{e_{p,i}}$ denotes the weight of the $i$-th edge in filament
$p$. The pairwise filament roughness is the average absolute value
of the difference between weights of adjacent edges. It reflects the
consistency of the edge weights along a filament which is typically
smaller within than across filaments (but cf.~Discussion). Moreover,
if the path consists of a single edge we take its weight as a roughness
measure. This choice increases the flexibility of the obtainable filament
covers and is necessary to avoid a cover by only individual edges
which contribute zero weight when weighted only according the first
line in Eq.~\ref{eq:supp_r_pair}. Another measures for the quality
of a filament is the all-to-all filament roughness (cf.~Eq.~\ref{eq:r_all})
\begin{eqnarray}
r_{p,\mathrm{all}} & = & \begin{cases}
\left(P-1\right)^{-1}\underset{i,j\in\left\{ 1,\dots,P\right\} }{\mathrm{max}}\left|w_{e_{p,i}}-w_{e_{p,j}}\right| & ,\,P>1\\
w_{e_{p,1}} & ,\,P=1
\end{cases},\,\,\,\,\,\,\,\,\,\label{eq:supp_r_all}
\end{eqnarray}
which is the average maximal difference between any edge weights in
a path $p$, and again the original weight of the edge is used for
a path of length one. Taking into account that most filaments are
only moderately bent, we may further wish to minimise the maximal
filament deflection angle between adjacent edges of a path $p$ (cf.~Eq.~\ref{eq:r_angle}),
\begin{eqnarray}
r_{p,\mathrm{angle}} & = & \underset{i\in\left\{ 1,\dots,P-1\right\} }{\mathrm{max}}\label{eq:supp_r_angle}\\
 &  & \,\,\,\mathrm{angle}\Biggl(v_{e_{p,i+1,1}}-v_{e_{p,i+1,0}},v_{e_{p,i,1}}-v_{e_{p,i,0}}\Biggr)\nonumber 
\end{eqnarray}
where $v_{e_{p,i,0}}$ and $v_{e_{p,i,1}}$ denote the positions of
the start and end nodes of the $i$-th edge of filament $p$, respectively.
Moreover, $\mathrm{angle}\left(v,v^{'}\right):=\arccos\left(\frac{v\cdot v^{'}}{\sqrt{v\cdot v}\sqrt{v^{'}\cdot v^{'}}}\right)$
is the Euclidean angle of two vectors $v$ and $v^{'}$ and $r_{p,\mathrm{angle}}=0^{\circ}$
corresponds to perfectly straight alignment.

\section{\label{sec:app_proof}Supplemental Material S2: Computational intractability
of the filament cover problem}

\begin{figure}
\begin{centering}
\includegraphics[width=1\textwidth]{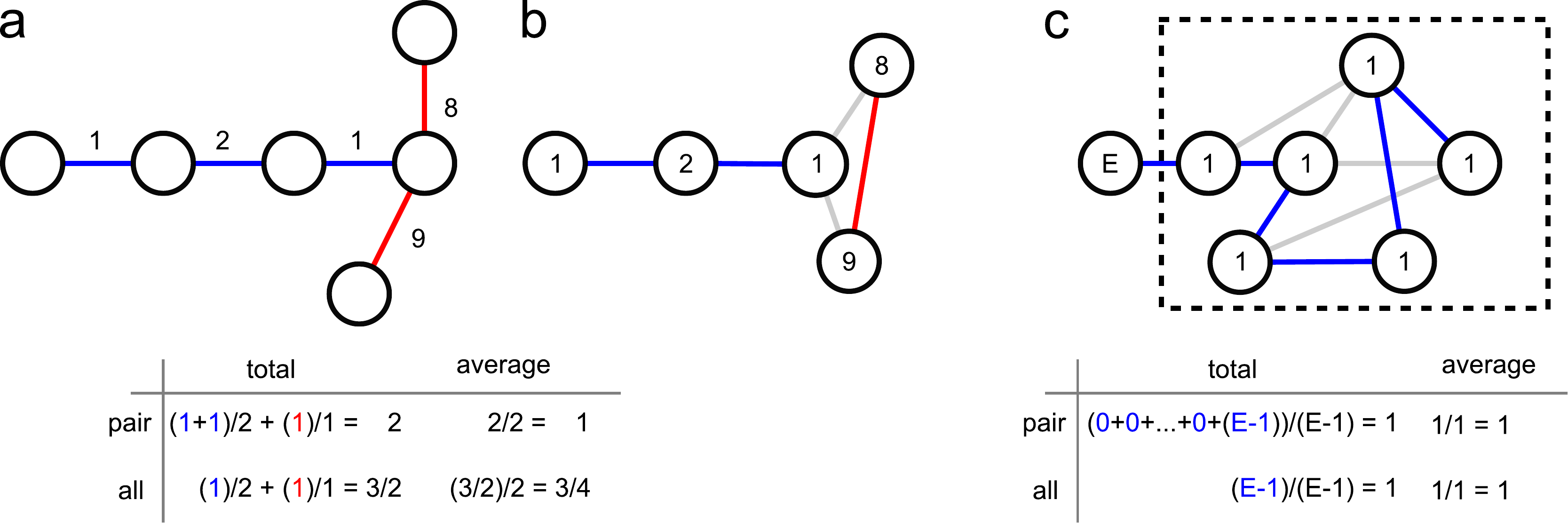}
\par\end{centering}

\caption{\textbf{\label{fig:app_proof}Proof of NP-hardness of the filament
cover problem.} \textbf{(a)} Optimal filament cover of an exemplary
(edge-weighted) graph. Table with cover roughnesses $R$ for minimisation
of total or average roughness and pairwise or all-to-all filament
roughness measure, respectively. \textbf{(b)} Corresponding (node-weighted)
line graph with equivalent path cover and the same roughness results
as for the (edge-weighted) graph in (a). \textbf{(c)} Extension of
an arbitrary graph with node weights $1$ by a node of weight $E$.
Here, finding a (node-weighted) path cover of roughness $R=\left(0+0+\dots+\left(E-1\right)\right)/\left(E-1\right)=1$
or less is equivalent to finding a Hamiltonian path. This equivalence
holds for covers minimising the total or average roughness of the
cover and using the pairwise or all-to-all filament roughness measure
(Eqs.~\ref{eq:supp_r_pair} and \ref{eq:supp_r_all}), see table.}
\end{figure}

The FCP is difficult to solve. This is intuitively clear as the number
of paths (let alone the number of path covers) increases rapidly with
the number of nodes $N$. Even in planar graphs, the number of closed
paths visiting each node once was shown to increase at least exponentially
with $N$ \cite{Buchin2007,Biswas2012}. We show now that the FCP
is NP-hard, even for planar, cubic graphs. Planar graphs can be drawn
on a plane without crossing edges. They are of particular relevance
since graphs that are generated from two-dimensional image data are
planar by construction \cite{Baumgarten2012,Obara2012a}. Cubic graphs
have only nodes of degree three. A proof of NP-hardness of a problem
for planar, cubic graphs directly implies its NP-hardness on general
graphs. The basic idea of a typical proof of computational complexity
is as follows \cite{Garey1979}: A problem of known complexity is
selected. By providing a constructive transformation or reduction,
a bijection between the known problem and the problem in question
is established, i.e., any yes-instance of the decision-version of
the known problem is mapped to a yes-instance of the decision-version
of the problem of interest and analogously for the no-instances. This
reduction proves that the two problems fall into the same class of
computational complexity. Our proof is by reduction from the Hamiltonian
path problem (HPP) on planar, cubic graphs which is known to be NP-complete
\cite{Garey1976}. The HPP asks, for a given graph, whether there
is a node-path which visits each node exactly once.

First, we note that finding a filament cover on an edge-weighted graph
$G$ is equivalent to finding a node-path cover on its node-weighted
line graph $L\left(G\right)$ (Fig.~\ref{fig:app_proof}A and B).
The line graph $L\left(G\right)$ of a graph $G$ has a node of weight
$w_{e}$ for each edge $e$ in $G$ and edges connecting two nodes
if the corresponding edges share a node in $G$.

Second, for a given line graph $L\left(G\right)$, we construct a
graph such that finding a node-path cover of weight $1$ or less is
equivalent to solving the HPP. To that end, we add one edge with a
terminal node to the line graph and set all original node-weights
to $1$ and the new node-weight to $E$ (Fig.~\ref{fig:app_proof}C).
Then, only a Hamiltonian path ensures a minimal weight of $R=C^{-A}\sum_{i=1}^{C}r_{p_{i}}=\frac{1}{1}\frac{\left(0+\dots+\left(E-1\right)\right)}{\left(E-1\right)}=1$,
for both pairwise and all-to-all filament roughness $r_{p}=\left\{ r_{p,\mathrm{pair}},r_{p,\mathrm{all}}\right\} $
(cf.~Eqs.~\ref{eq:supp_r_pair} and \ref{eq:supp_r_all}) and both
minimisation of total and average filament roughness, i.e., $A\in\left\{ 0,1\right\} $.

Finally, we show that finding a Hamiltonian path on a line graph of
a planar, cubic graph is NP-complete. It was shown that the HPP is
NP-complete on general line graphs via a reduction from the HPP in
cubic graphs \cite{Bertossi1981}. This reduction remains valid when
planar, cubic graphs are used instead of cubic graphs, for which NP-completeness
of HPP is known \cite{Garey1976}. Therefore, the decision version
of the FCP is NP-complete and the FCP is NP-hard, as claimed. Since
the FCP is NP-hard on planar, cubic graphs, it is (at least) NP-hard
on general graphs.

\section{\label{sec:app_poly}Supplemental Material S3: The filament cover
problem on trees is solvable in polynomial time }

While we showed that the FCP is NP-hard on general and even planar,
cubic graphs, it is solvable in polynomial time on trees. The polynomial
algorithm outlined here is similar to those proposed to find an unrestricted
node-path cover where each vertex may be included in multiple paths
of minimum cardinality or minimum weight \cite{Lin2006}.

The basic idea is to assume that a certain path covering a certain
edge is in the cover (in a tree, there are at most $N\left(N-1\right)/2=\mathcal{O}\left(N^{2}\right)$
paths to choose from). Upon removal, the tree is split into potentially
multiple forests (at most $\mathcal{O}\left(N\right)$), each tree
of which is decomposed in the same way. The procedure is repeated
for each edge (clearly $\mathcal{O}\left(N\right)$ in a tree). Thus,
this results in a dynamic programming algorithm which has an overall
polynomial time complexity of $\mathcal{O}\left(N^{4}\right)$.

The above procedure assumes non-overlapping paths and may be extended
to limitedly overlapping paths. For the completely unrestricted case,
there would be $\mathcal{O}\left(2^{\mathrm{\#paths}}\right)=\mathcal{O}\left(2^{N^{2}}\right)$
combinations for covering a given edge to chose from in the first
step, and the time complexity of the algorithm would be exponential.
However, the problem remains polynomial if we allow only $k$-fold
overlaps, $k=\mathcal{O}\left(1\right)$, i.e., each edge may be covered
by at most $k$ paths. In the first step of the above algorithm, a
given edge may then be covered by at most $\mathcal{O}\left(\left(\begin{array}{c}
N\left(N-1\right)/2\\
k
\end{array}\right)^{\,}\right)=\mathcal{O}\left(N^{2k}\right)$ edges and consequently the time complexity of the full algorithm
is $\mathcal{O}\left(N^{2k+2}\right)$.

\section{\label{sec:app_setcover}Supplemental Material S4: Approximation
algorithm for the filament cover problem}

Since the FCP is NP-hard even on planar, cubic graphs, we need suitable
approximation algorithms. In particular, the approximation algorithms
should allow overlapping filaments as well as looped filaments. A
natural choice seems to be the formulation of the FCP as a set cover
problem (SCP) \cite{Karp1972}: 
\begin{quote}
Given an object set $\mathcal{U}$, called universe, and a set $\mathcal{S}$
of sets with costs $c_{s}$, $s\in\mathcal{S}$:

Find a subset $\mathcal{S}_{\mathrm{set}}\subseteq\mathcal{S}$ with
minimal total (or average) cost such that each element in $\mathcal{U}$
is covered (at least) once. 
\end{quote}
In our case, the universe corresponds to the set of edges of the given
graph ($\mathcal{U}\hat{=}\mathcal{E}$), a set corresponds to a path
($s\hat{=}p$), the cost of a set corresponds to the roughness of
a path ($c_{s}\hat{=}r_{p}$), and the set cover corresponds to the
desired filament cover ($\mathcal{S}_{\mathrm{set}}\hat{=}\mathcal{P}_{\mathrm{fil}}$).
We note, that this formulation of the SCP allows overlapping sets,
$s\cap s^{'}\neq\emptyset$, $s,s^{'}\in\mathcal{S}$, which directly
translates into overlapping filaments in our FCP. By requiring that
each element in $\mathcal{U}$ is contained in $\mathcal{S}_{\mathrm{set}}$
exactly once, we may exclude filament overlaps.

An open task is then the generation of a suitable set of paths ($\mathcal{S}\hat{=}\mathcal{P}$).
Since for a general graph it is not feasible to find all paths $\mathcal{P}$
(cf.~the motivation of the NP-hardness proof of the FCP above), we
need to find a representative subset ,$\mathcal{P}^{'}$, of paths.
We propose two approaches: (1) We sample paths from $T=100$ random
minimal spanning trees (RMST) of $G$. To obtain a RMST, each edge
is assigned a uniformly distributed random weight and the minimum
spanning tree with respect to these weights is computed. Each tree
has $N\left(N-1\right)/2$ non-trivial, undirected paths that we add
to our set $\mathcal{P}^{'}$. However, the paths in a tree cannot
contain loops. (2) We perform a modified breadth-first search (BFS)
on the nodes, store the generated paths, and stop the search for a
path $p$ when it violates a straightness criterion, e.g., $r_{p,\mathrm{angle}}<60^{\circ}$
(cf.~Eq.~\ref{eq:supp_r_angle}) which is used throughout the paper.
We add all permitted paths to $\mathcal{P}^{'}$. We note that for
all real-world filamentous graphs, due to filament thickness, there
are spatial constraints on the number of nodes of a graph as well
as on the node degrees. Moreover, for the filamentous networks considered
here, the radius of curvature of a filament is typically not much
smaller than the region of interest. The number of loops is further
reduced by the straightness criterion which eliminates paths with
a small radius of curvature. Hence, the number of loops in the network
is restricted and our heuristically modified BFS allows for loops
and yields a representative set $\mathcal{P}^{'}$ in reasonable time.

The SCP may be expressed as a binary fractional linear program \cite{Vazirani2001},
and we analogously write the FCP as 
\begin{eqnarray}
\mathrm{minimize} & \, & \frac{\sum_{p\in\mathcal{P}}r_{p,\mathrm{pair}}x_{p}}{\left(\sum_{p\in\mathcal{P}}x_{p}\right)^{A}}\label{eq:supp_lp_frac}\\
\mathrm{subject\,to} & \, & \sum_{p:e\in p}x_{p}\geq1\,\mathrm{for\,all\,}e\in\mathcal{E}\nonumber \\
 & \, & x_{p}\in\left\{ 0,1\right\} \,\mathrm{for\,all\,}p\in\mathcal{P}^{'},\nonumber 
\end{eqnarray}
where in the first line $A\in\left\{ 0,1\right\} $ determines whether
the total or the average roughness is minimised. In the second line,
equality holds for an exact cover. For $A=0$, Eq.~\ref{eq:supp_lp_frac}
is a binary linear program that may be solved using well-established
and efficient algorithms \cite{Schrijver1998,Linderoth2005}. 

For $A=1$, the fractional problem may be rewritten as a binary linear
program as well \cite{Wu1997,Yue2013}. To that end, we introduce
new variables $y=\left(\sum_{p\in\mathcal{P}}x_{p}\right)^{-1}$ and
$z_{p}=x_{p}y$, $p\in\mathcal{P}^{'}$. The latter expression is
non-linear but may be replaced by a set of binary linear equations,
yielding 
\begin{eqnarray}
\mathrm{minimize} & \, & \sum_{p\in\mathcal{P}}r_{p}z_{p}\label{eq:supp_lp_lin}\\
\mathrm{subject\,to} & \, & \sum_{p:e\in p}z_{p}\geq y\,\mathrm{for\,all\,}e\in\mathcal{E}\nonumber \\
 & \, & \sum_{p\in\mathcal{P}}z_{p}=1\nonumber \\
 & \, & y\geq0\nonumber \\
 & \, & y-z_{p}\leq M-Mx_{p}\nonumber \\
 & \, & z_{p}\leq y\nonumber \\
 & \, & z_{p}\leq Mx_{p}\nonumber \\
 & \, & z_{p}\geq0\nonumber \\
 & \, & x_{p}\in\left\{ 0,1\right\} \,\mathrm{for\,all\,}p\in\mathcal{P}.\nonumber 
\end{eqnarray}
Here, $M$ is a sufficiently large constant that needs to exceed any
$y$ (cf.~the Big $M$ method \cite{Griva2009}). Since $y=\left(\sum_{p\in\mathcal{P}^{'}}x_{p}\right)^{-1}\leq1$
for the cover of any non-empty graph, we choose $M=2$. 

Thus, there are a number of options in our FCP: The input set of paths
may be obtained by using a modified BFS or from sampling RMSTs or
(denoted by either \emph{BFS} or \emph{RMST}). The filaments may overlap
or not (\emph{over}/\emph{exact}). The objective of the FCP may be
the minimisation of the total or the average roughness (\emph{total}/\emph{avg}).
The roughness of a filament may be measured by the pairwise or the
all-to-all filament roughness (\emph{pair}/\emph{all}). Solutions
of the FCP with different options are compared in the Results. 

An implementation of the presented approximation schemes to the FCP
with the described options is supplied as an open-source tool, ``DeFiNe''
(\textbf{De}composing \textbf{Fi}lamentous \textbf{Ne}tworks), under
GLP3 at \href{http://mathbiol.mpimp-golm.mpg.de/DeFiNe/}{http://mathbiol.mpimp-golm.mpg.de/DeFiNe/}.
DeFiNe is programmed in Python \cite{VanRossum2011} and employs the
packages SciPy \cite{Olivier2002}, NetworkX \cite{Hagberg2008},
and cvxopt \cite{Dahl2006} and PyGTK \cite{Finlay2005} for a simple
and user-friendly graphical user interface. DeFiNe takes as input
a weighted graph in the standard~.gml file format \cite{Himsolt1997}
and outputs a standard~.gml graph with filament identities stored
as edge colours. Node coordinates may be included in the input file
to enable the modified BFS that takes into account edge alignments.
Furthermore, manual filament assignments may be included in the input
file and the similarity with the automatically obtained filament cover
is assessed as described below. In addition, DeFiNe generates a standard,
human-readable~.csv-table of various individual filament measures
for custom analyses. The filamentous structure as well as the manual
filament assignments shown in Fig.~\ref{fig:res_artificial} are
available as a~.gml file under the above internet address for demonstration
purposes.

\section{\label{sec:app_extraction}Supplemental Material S5: Extraction of
weighted networks from images}

The procedure used to extract weighted networks from image data is
similar to those proposed in \cite{Baumgarten2012,Obara2012a}: (1)
The original grey-scale image are pre-processed to enhance the filamentous
structures. Here, a vesselness filter with kernel width of $2\,\mathrm{pixels}$
was used for simplicity \cite{Frangi1998}. (2) In the filtered image,
the filamentous structures are separated from the background by applying
an adaptive median threshold with a block size of $49\,\mathrm{pixels}$,
whereby moderate variations of this size leave our findings largely
unchanged. (3) The resultant binary image is skeletonised to obtain
the filament centre lines \cite{Haralick1987}. (4) Then, the nodes
of the network under construction are extracted as terminal points,
branching points, or crossings of skeleton branches. (5) An edge is
inserted between two nodes if they are directly connected via the
skeleton. (6) Finally, the edges are weighted by integrating the intensity
of the underlying original grey-scale image smoothed with a Gaussian
filter with a standard deviation of $5\,\mathrm{pixels}$ along the
filament and taking its average per unit length of the filament. For
the images of the simulated galaxy clusters, the structures obtained
by a model-based filter from \cite{Stoica2005}, Figure 6 left middle
and bottom rows, are directly employed as binary images and the networks
representations are obtained as described above.

\section{\label{sec:app_similarity}Supplemental Material S6: Quality assessment
of filament covers via structure-aware partition similarity measures}

\begin{figure}
\centering{}\includegraphics[width=0.9\textwidth]{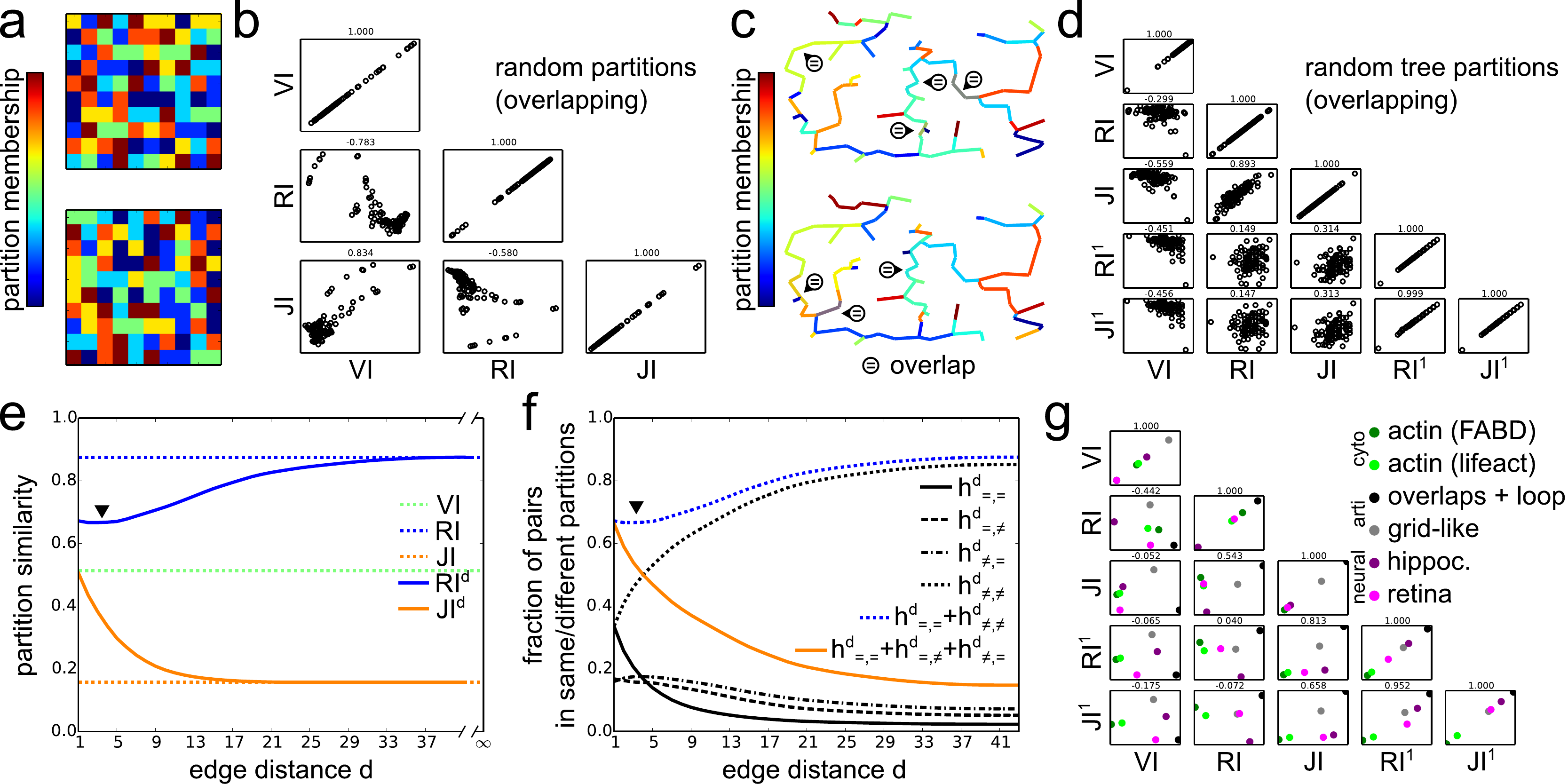}
\end{figure}

\begin{figure}
\caption{\textbf{\label{fig:app_similarity}Comparison of classical and extended
partition similarity measures.} Analysis of $100\times2$ random partitions
of sets of $100$ numbers plus $10$ duplicate ones into $5-10$ partitions
(a-b). Analysis of a $100\times2$ path covers of Euclidean minimum
spanning trees with $100$ nodes distributed uniformly in the unit
square, where the paths are drawn randomly and added if the overall
overlap of paths is below $10$ edges (c-f). Analysis of the similarities
between the manual and automated decompositions of the networks studied
in the paper (g). \textbf{(a)} Colour-representation of two exemplary
random partitions as explained above. \textbf{(b)} The classical partition
similarity measures $\mathrm{VI}$, $\mathrm{RI}$, and $\mathrm{JI}$
are not correlated (cf.~Kendall rank correlation coefficients $\tau<0.9$)
and may lead to opposing conclusions for the similarity of different
partitions. \textbf{(c)} Two exemplary random tree path covers with
overlaps ({\scriptsize{}\textcircled{=}}). \textbf{(d)} The classical
partition similarity measures $\mathrm{VI}$, $\mathrm{RI}$, and
$\mathrm{JI}$ show no correlation among themselves (except for the
pairing of $\mathrm{RI}$ and $\mathrm{JI}$), nor with the structure-aware
$\mathrm{RI}^{1}$ and $\mathrm{JI}^{1}$. In contrast, $\mathrm{RI}^{1}$
and $\mathrm{JI}^{1}$ are very strongly correlated (cf.~$\tau>0.9$)
and yield consistent results for the similarity of different partitions.
\textbf{(e)} Similarity of the partitions shown in (c) in dependence
on maximal distance $d$ between considered pairs of edges (cf.~Fig.~\ref{fig:res_actin}
for a detailed discussion). The $\mathrm{RI}^{d}$ shows a non-monotonous
dependency on $d$ (triangle). \textbf{(f)} This non-monotonicity
of $\mathrm{RI}^{d}$ may be explained by the entries of the contingency
table $h_{\times,\times^{'}}$, $\times,\times^{'}\in\left\{ =,\protect\neq\right\} $.
For small distances $d$, the fraction of true positives (solid black,
$h_{=,=}$) drops slower than the fraction of true negatives (dotted
black, $h_{\protect\neq,\protect\neq}$) and for larger $d$, this
trend is reversed. Hence, their sum (dashed blue) shows a minimum
at intermediate distances $d$ (triangle). In contrast, when summed
up (solid yellow), the fast drop in the fraction of true positives
dominates over the slightly non-monotony of the false positives and
negatives. \textbf{(g)} For the investigated artificial and biological
networks, the classical measures $\mathrm{VI}$, $\mathrm{RI}$, and
$\mathrm{JI}$ yield partially opposing results on the similarity
of the manual assignment and the automated decomposition (cf.~$\tau<0$).
The structure-aware similarity measures $\mathrm{RI}^{1}$ and $\mathrm{JI}^{1}$
are strongly correlated and yield consistent results (cf.~$\tau=0.952$).}
\end{figure}

The extraction of the filamentous networks from image data enables
comparison of the automated filament cover with manual filament assignments.
Both, automated cover and manual assignment may be regarded as partitions
(where we allow overlapping subsets as well). As measures for the
similarity of the automated and manual partitions we use the variation
of information,$\mathrm{VI}$, the Jaccard index, $\mathrm{JI}$,
and the Rand index, $\mathrm{RI}$, which are commonly used and were
shown to estimate similarity reliably for distant and close partitions
alike \cite{Saporta2002,Meila2005,Deneud2006}. For given partitions
$\mathcal{C}=\left\{ C_{1},\dots,\mathcal{C}_{C}\right\} $ and $\mathcal{C}^{'}=\left\{ C_{1}^{'},\dots,\mathcal{C}_{C^{'}}^{'}\right\} $,
they are computed via
\begin{eqnarray}
\mathrm{VI}\left(\mathcal{C},\mathcal{C}^{'}\right) & = & 1+\left(U\log U\right)^{-1}\cdot\label{eq:supp_VI}\\
 &  & \,\,\,\,\,\,\,\:\cdot\sum_{i,j}g_{i,j}\left(\log\left(\frac{g_{i,j}}{g_{\cdot,j}}\right)+\log\left(\frac{g_{i,j}}{g_{i,\cdot}}\right)\right),\nonumber \\
\mathrm{RI}\left(\mathcal{C},\mathcal{C}^{'}\right) & = & \frac{h_{=,=}+h_{\neq,\neq}}{h_{=,=}+h_{=,\neq}+h_{\neq,=}+h_{\neq,\neq}},\label{eq:supp_RI}\\
\mathrm{JI}\left(\mathcal{C},\mathcal{C}^{'}\right) & = & \frac{h_{=,=}}{h_{=,=}+h_{=,\neq}+h_{\neq,=}},\label{eq:supp_JI}
\end{eqnarray}
where $U=\sum_{i=1}^{C}\left|\mathcal{C}_{i}\right|=\sum_{j=1}^{C^{'}}\left|\mathcal{C}_{j}^{'}\right|$,
$g_{i,j}=\left|\mathcal{C}_{i}\cap\mathcal{C}_{j}^{'}\right|$, $g_{\cdot,j}=\sum_{i=1}^{C}g_{i,j}$,
and $g_{i,\cdot}=\sum_{j=1}^{C^{'}}g_{i,j}$. The contingency tables
$h_{\times,\times^{'}}$, $\times,\times^{'}\in\left\{ =,\neq\right\} $,
provide the numbers of edge pairs which are in the same or different
sets in the two partitions, respectively, and is related to $g_{i,j}$
as shown in \cite{Hubert1985}. All measures are restricted to the
unit interval with larger values reflecting higher similarity \cite{Meila2003}.

While these measures of partition similarity are widely used \cite{Meila2005,Lancichinetti2009},
they pose some difficulties. The variation of information, $\mathrm{VI}$,
is only well-defined for disjoint partitions, which occur for non-overlapping
filaments. While the Jaccard index, $\mathrm{JI}$, and the Rand index,
$\mathrm{RI}$, cover intersecting partitions they may generally yield
opposing results. We demonstrate this inconsistency by investigating
two types of random partitionings: First, for $100$ repetitions,
we randomly partitioned $2$ sets of $100$ numbers and up to $10$
duplicates (to simulate overlapping filaments) into $5-10$ random
partitions (Fig.~\ref{fig:app_similarity}a). While $\mathrm{VI}$
and $\mathrm{JI}$ were correlated (Fig.~\ref{fig:app_similarity}b;
cf.~Kendall rank correlation coefficient $\tau>0$), the other two
combinations showed a strong negative correlation (cf.~$\tau<0$).
Second, to study filament covers that resemble the decomposition of
real filamentous networks more closely, we constructed a relative
neighbourhood graph \cite{Toussaint1980,Supowit1983} with $100$
nodes uniformly distributed in the unit square and computed a random
minimum spanning tree (Fig.~\ref{fig:app_similarity}c). For $100$
repetitions of this procedure, we partitioned the resultant tree into
filaments by choosing a path at random and adding it to the decomposition
if the total overlap of any two paths already in the decomposition
is below $10$ edges (cf.~{\scriptsize{}\textcircled{=}} for overlaps).
Again, the correlation among the classical similarity measures was
poor or negative (Fig.~\ref{fig:app_similarity}d; except for the
correlation between $\mathrm{RI}$ and $\mathrm{JI}$; $\left|\tau\right|<0.6$).
Although other measure for the similarity of intersecting partitions
have been proposed \cite{Goldberg2010,Lancichinetti2009a,Lancichinetti2009b},
we adhere to $\mathrm{RI}$ and $\mathrm{JI}$ for simplicity.

More severely, however, the above similarity measures do not take
into account the structure of the graph $G$ underlying the (edge-)partitions
induced by the obtained filament covers. To date, we are only aware
of structure-aware similarity measures for the comparison of partitions
whose items are distributed in Euclidean space \cite{Zhou2005,Bae2010,Coen2010}.
Yet, these approaches do not take into account the explicit graph
structure of the partitions. To remedy this shortcoming, we introduce
a suite of measures, the structure-aware Rand and Jaccard index, $\mathrm{RI}^{d}$
and $\mathrm{JI}^{d}$, respectively. To that end, the contingency
tables $h_{\times,\times^{'}}$ in Eqs.~\ref{eq:supp_RI} and \ref{eq:supp_JI}
are replace by distance dependent $h_{\times,\times^{'}}^{d}$,
\begin{eqnarray}
\mathrm{RI}^{d}\left(\mathcal{C},\mathcal{C}^{'}\right) & = & \frac{h_{=,=}^{d}+h_{\neq,\neq}^{d}}{h_{=,=}^{d}+h_{=,\neq}^{d}+h_{\neq,=}^{d}+h_{\neq,\neq}^{d}},\label{eq:supp_RId}\\
\mathrm{JI}^{d}\left(\mathcal{C},\mathcal{C}^{'}\right) & = & \frac{h_{=,=}^{d}}{h_{=,=}^{d}+h_{=,\neq}^{d}+h_{\neq,=}^{d}}\label{eq:supp_JId}
\end{eqnarray}
where $h_{\times,\times^{'}}^{d}$, $\times,\times^{'}\in\left\{ =,\neq\right\} $,
$d\in\mathbb{N}_{>0}$, count the number of edge pairs which are in
the same or different sets in the two partitions, respectively, and
which are separated by at most $d$ nodes in $G$. More precisely,
we define 
\begin{eqnarray}
h_{\times,\times^{'}}^{d} & = & \Biggl\{\mathrm{\#}\left(e_{0},e_{1}\right)\mid e_{0}\in\mathcal{C}_{i}\cap\mathcal{C}_{i^{'}}^{'},\,e_{1}\in\mathcal{C}_{j}\cap\mathcal{C}_{j^{'}}^{'},\label{eq:supp_hdis}\\
 &  & \mathrm{\,with\,}i\times j\mathrm{\,and\,}i^{'}\times^{'}j^{'}\mathrm{\,and\,}D_{L\left(G\right)}\left(e_{0},e_{1}\right)\leq d\Biggr\},\nonumber 
\end{eqnarray}
where $\#\left(e_{0},e_{1}\right)$ is the number of edges $\left(e_{0},e_{1}\right)$
and $D_{L\left(G\right)}\left(e_{0},e_{1}\right)$ is the length of
the shortest path between nodes in the line graph $L\left(G\right)$
of $G$ corresponding to the edges $e_{0}$ and $e_{1}$. For example,
$h_{=,=}^{0}$ counts the the number of adjacent edges which are in
the same set in both partitions (local perspective). In contrast,
$h_{\times,\times^{'}}^{\infty}\equiv h_{\times,\times^{'}}$ reproduce
the original measures which do not take into account the positions
of edges in the graph (global perspective). 

To investigate the performance of our extended, structure-aware partition
similarity measures, $\mathrm{RI}^{d}$ and $\mathrm{JI}^{d}$, we
apply them to the artificial graph-based random partitions described
above (cf.~Fig.~\ref{fig:app_similarity}c). Indeed, when considering
the partition membership of neighbouring edges only, i.e., $\mathrm{RI}^{1}$
and $\mathrm{JI}^{1}$, the similarity measures yield very consistent
results (Fig.~\ref{fig:app_similarity}d; cf.~$\tau=0.999$) in
contrast to the lower correlation with the classical similarity measures
(cf.~$\tau<0.9$). Investigating the dependency of $\mathrm{RI}^{d}$
and $\mathrm{JI}^{d}$ on the distance $d$ for the tree filament
covers shown in Fig.~\ref{fig:app_similarity}d, we find that $\mathrm{RI}$
and $\mathrm{JI}$ (Fig.~\ref{fig:app_similarity}e; dotted blue
and yellow) over- and underestimate the partition similarity with
respect to $\mathrm{RI}^{1}$ and $\mathrm{JI}^{1}$ (Fig.~\ref{fig:app_similarity}e;
solid blue and yellow). Furthermore, we find that $\mathrm{RI}^{d}$
is non-monotonic in $d$ (Fig.~\ref{fig:app_similarity}e; cf.~the
black triangle). These errors in estimation are explained by the large
fraction of false negatives ($h_{\neq,\neq}$) and the small fraction
of true positives ($h_{=,=}$), respectively, which dominate for large
distances $d$, i.e., the limit in which the graph structure is ignored
(Fig.~\ref{fig:app_similarity}f; dotted black and solid black).
Due to the differential increase/decrease of $h_{\neq,\neq}$/$h_{=,=}$,
their combination and therefore $\mathrm{RI}^{1}$ is non-monotonic
(Fig.~\ref{fig:app_similarity}f; dashed blue). Finally, we observe
opposing results of the classical partition similarity measures also
for the filament covers of artificial and biological filamentous networks
investigated in the main text, while our extended, structure-aware
measures $\mathrm{RI}^{1}$ and $\mathrm{JI}^{1}$ provide consistent
similarity results (Fig.~\ref{fig:app_similarity}g).

\section{\label{sec:app_robust}Supplemental Material S7: Robustness of filament
covers against incomplete knowledge of underlying network structure
and image noise}

\begin{figure}
\centering{}\includegraphics[width=1\textwidth]{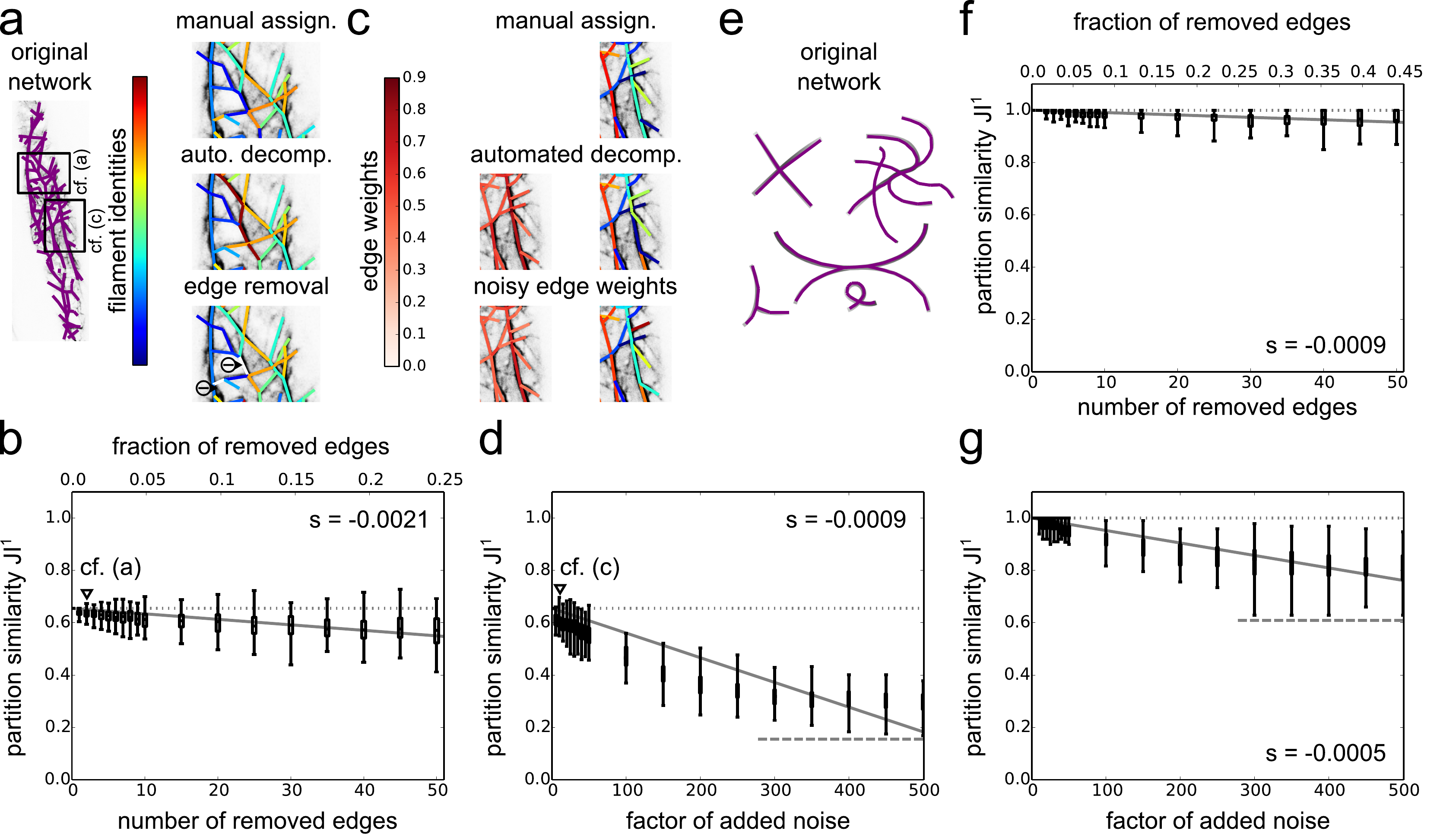}
\end{figure}

\begin{figure}
\caption{\textbf{\label{fig:app_robust}Analyses of robustness of filament
covers against incomplete knowledge of network and image noise.} A
cytoskeletal and a contrived network are decomposed automatically
by solving the FCP with options given in Fig.~\ref{fig:res_actin}c
and Fig.~\ref{fig:res_artificial}c, respectively.\textbf{ (a)} Overlay
of extracted actin network structure and original image data (left
panel). Sections of cytoskeletal network with edge colours representing
the manual assignment, the optimal filament cover obtained for the
full, non-disrupted network, and the optimal filament cover after
removal of two edges which are shown in white (right panels). \textbf{(b)}
Similarity of manual filament assignment and automated filament covers
after removal of increasing numbers of edges, measured by structure-aware
Jaccard index $\mathrm{JI}^{1}$. On average, $\mathrm{JI}^{1}$ decreases
with the number of removed edges as shown by a linear fit with slope
$s=-0.0021$ (solid grey line). Occasionally, the removal of edges
increases the accuracy of the filament cover above the accuracy of
the original solution (dotted grey line and triangle; cf.~panel (a)).
\textbf{(c)} Sections of cytoskeletal network with edge colours representing
the original edge weights and the edge weights after adding Gaussian
noise (left panels). Sections of cytoskeletal network with edge colours
representing the manual assignment, the optimal filament cover obtained
for the full, non-disrupted network, and the optimal filament cover
after adding Gaussian noise (right panels). \textbf{(d)} Similarity
$\mathrm{JI}^{1}$ of manual filament assignment and automated filament
covers after adding Gaussian noise. On average, $\mathrm{JI}^{1}$
decreases with increasing noise factor as shown by a linear fit with
slope $s=-0.0009$ (solid grey line). Occasionally, the noisy edge
weights lead to an increase in accuracy of the filament cover above
the accuracy of the original solution (dotted grey line and triangle;
cf.~panel (c)). The decrease levels off for large noise factors and
$\mathrm{JI}^{1}$ approaches a constant value (dashed grey line).
\textbf{(e)} Overlay of extracted contrived network structure and
original image data. \textbf{(f)} Results for the contrived network
analogue to those presented for the cytoskeletal network in panel
(b). The average change in $\mathrm{JI}^{1}$ per removed edge is
captured by a linear fit with slope $s=-0.0009$. \textbf{(g)} Results
for the contrived network analogue to those presented for the cytoskeletal
network in panel (f). The average change in $\mathrm{JI}^{1}$ per
unit increase in the noise factor is captured by a linear fit with
slope $s=-0.0005$.}
\end{figure}

Our approach enables accurate decomposition of a given filamentous
network into its constitutive filaments (cf.~Results). However, the
preceding extraction of the network from image data is often non-trivial
(cf.~Methods). Therefore, to assess the robustness of our approach,
we test how the accuracy of our filament decomposition is affected
(1) by incomplete knowledge of the true underlying network structure
and (2) by image noise which affects the edge weights of the extracted
network. We perform these analyses for the actin cytoskeleton shown
in Fig.~\ref{fig:res_actin} (Fig.~\ref{fig:app_robust}a, left
panel) and the contrived network shown in Fig.~\ref{fig:res_artificial}
(Fig.~\ref{fig:app_robust}e).

(1) First, we start from the original, weighted network and randomly
remove one of the $E$ edges to model erroneous segmentation. For
the disrupted network, we recompute the optimal filament cover (with
the same options as in Figs.~\ref{fig:res_artificial}c and \ref{fig:res_actin}c,
respectively) and calculate its agreement with the original manual
segmentation (measured by the structure-aware Jaccard index $\mathrm{JI}^{1}$;
the removed edge is assigned a dummy label). We repeat the procedure
for $E$ networks from which a single, randomly chosen edge has been
removed. Next, we repeat the procedure for $E$ networks from which
a randomly chosen double of edges has been removed. We then proceed
with triplets, quartets, and so on up to subsets of $50$ randomly
chosen edges.

As expected, the removal of increasing numbers of edges typically
decreases the agreement of the automated filament cover with the manual
assignment for the cytoskeletal as well as the contrived network (Fig.~\ref{fig:app_robust}b
and f). For both networks, however, the decrease is slow and $\mathrm{JI}^{1}$
increases only by around $0.002$ per removed edge (cf.~Fig.~\ref{fig:app_robust}b
and f, solid grey line indicates linear fit). Interestingly, for the
actin cytoskeleton, the removal of certain edges may even increase
the accuracy of the filament cover (Fig.~\ref{fig:app_robust}a,
right panels show manual filament assignment and automated filament
cover the original network, and an exemplary filament cover obtained
after the removal of two edges, coloured white here, which improves
the agreement with the manual assignment; cf.~Fig.~\ref{fig:app_robust}b,
dotted grey line and triangle). 

(2) Second, we simulate image noise by adding centred Gaussian noise
$\Delta w$ to the edge weights of the original network with
\begin{eqnarray}
\mathrm{E}\left[\Delta w\right] & = & 0,\label{eq:supp_noiseE}\\
\mathrm{\mathrm{Sd}\left[\Delta w\right]} & = & \left(1+\frac{f}{100}\right)w.\label{eq:supp_noiseS}
\end{eqnarray}
We normalise the standard deviation of the added noise by the original
edge weights to avoid extreme fluctuations, and $f$ is referred to
as noise factor. For each noise factor, we construct $100$ networks,
recompute the optimal filament covers, and measure their agreement
with the manual filament assignment, as in the first scenario above.

For both the contrived and the cytoskeletal network, the accuracy
of the filament cover decreases with increasing noise, as expected
(Fig.~\ref{fig:app_robust}d and g). However, this decrease in accuracy
is slow and $\mathrm{JI}^{1}$ decreases by less than $0.001$ when
increasing the standard deviation of the noise by $1\%$ of the original
edge weights, i.e., when increasing the noise factor by one (cf.~Fig.~\ref{fig:app_robust}d
and g, solid grey lines indicate linear fits). We note that with increasing
edge noise the accuracy of the filament cover approaches a constant,
non-zero $\mathrm{JI}^{1}$ which reflects that some information about
the filament structure maybe obtained from the topology of the network
alone, irrespective of the edge weights (cf.~Fig.~\ref{fig:app_robust}d
and g, dashed grey lines).

\section{\label{sec:app_movie}Supplemental Material S8: Filament analysis
for networks extracted from movie of plant actin cytoskeleton}

\begin{figure}
\centering{}\includegraphics[width=1\textwidth]{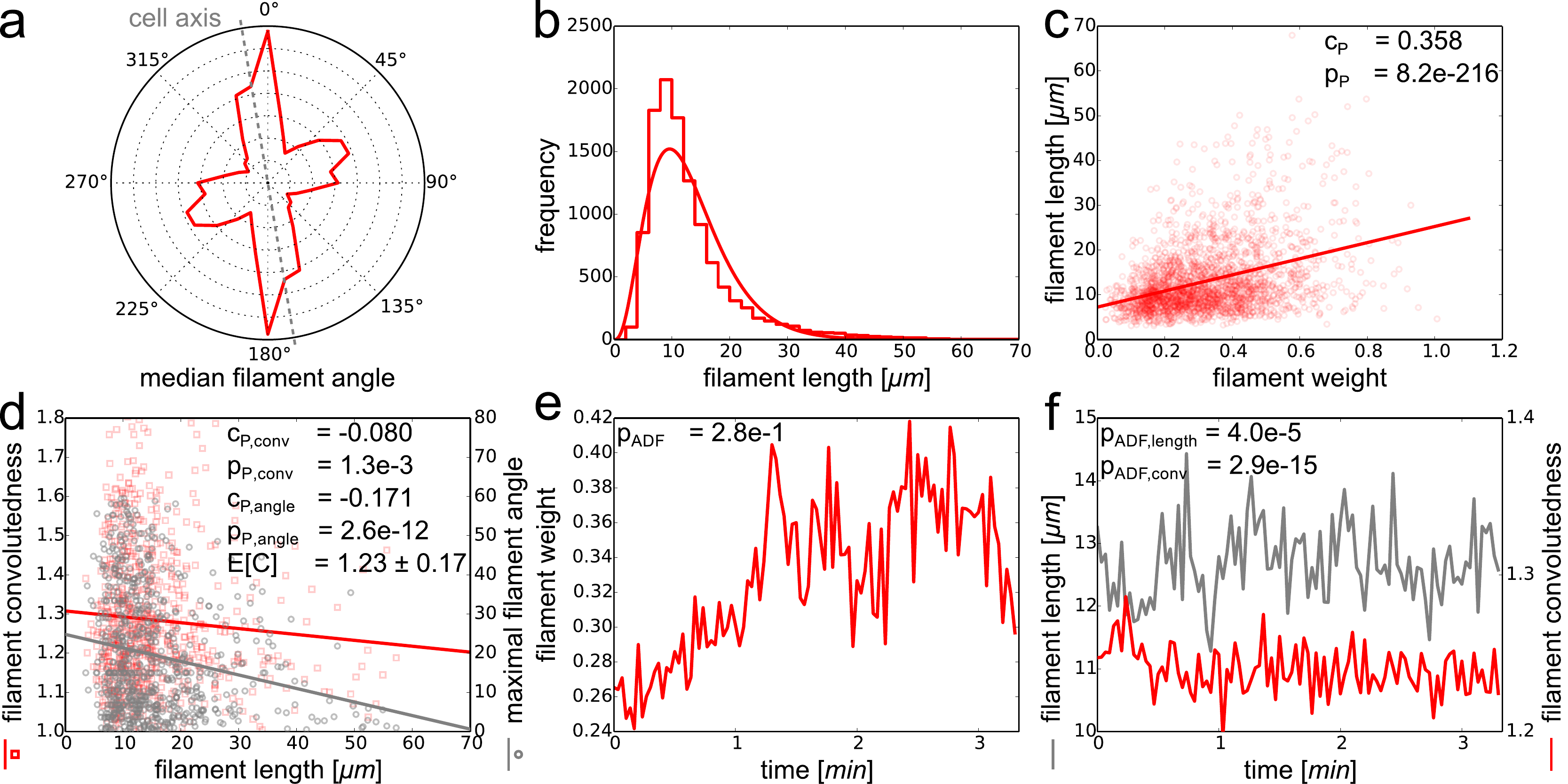}
\end{figure}

\begin{figure}
\caption{\textbf{\label{fig:app_movie}Filament analyses of $100$ cytoskeletal
networks.} Results from filament decompositions of $100$ cytoskeletal
networks extracted from a movie of a plant cytoskeleton over $200\,\mathrm{s}$.
The cytoskeletal networks are decomposed automatically by solving
the exact FCP (\emph{exact}) for paths from a modified breadth-first
search (\emph{BFS}) and by minimising the total (\emph{total}) pairwise
filament roughness (\emph{pair}; cf.~Fig.~\ref{fig:res_actin}).\textbf{
(a)} The distribution of median filament angles shows that the majority
of filaments is aligned parallel to the cell axis (grey dashed line).
\textbf{(b)} Filament lengths (bars) follows a gamma distribution
(line shows maximum likelihood fit). \textbf{(c)} Filament length
correlates with filament weight (cf.~linear regression and Pearson
correlation coefficient $c_{P}>0$ and $p$-value $p_{\mathrm{P}}<0.05$)
\textbf{(d)} Scatter plot of filament convolutedness versus filament
length shows a negative correlation (cf.~red squares, $c_{P,\mathrm{conv}}<0$,
and $p_{\mathrm{P},\mathrm{conv}}<0.05$) with an average convolutedness
of $\mathrm{E}\left[C\right]=1.23\pm0.17$. The maximum filament angle
correlates negatively with the filament length (cf.~grey circles,
$c_{P,\mathrm{angle}}<0$, and $p_{\mathrm{P},\mathrm{angle}}<0.05$),
indicating that longer (and thicker, cf.~(c)) filaments are less
curved. \textbf{(e)} Time series of average filament weight over $200\,\mathrm{s}$
shows large fluctuations and is non-stationary (cf.~augmented Dickey-Fuller
test $p$-value $p_{\mathrm{ADF}}\geq0.05$). \textbf{(f)} Time series
of filament length and convolutedness are stationary over the recording
period (cf.~$p_{\mathrm{ADF,length}}<0.05$ and $p_{\mathrm{ADF,conv}}<0.05$).}
\end{figure}

To further strengthen our statistical analyses of cytoskeletal actin
filaments (cf.~Fig.~\ref{fig:res_actin}), we investigate a complete
movie of a plant cytoskeleton of $100$ frames over $200\,\mathrm{s}$
(cf.~Methods for details). For each frame, we extract a weighted
network representation of the cytoskeleton as described above (cf.~Methods
for details) and solve the FCP with options described in Fig.~\ref{fig:res_actin},
i.e., we solve the exact FCP (\emph{exact}) for paths from a modified
breadth-first search (\emph{BFS}) and by minimising the total (\emph{total})
pairwise filament roughness (\emph{pair}). Analysis of various properties
of the automatically obtained filaments confirms our findings in Fig.~\ref{fig:res_actin}:
The filaments show a preferential alignment parallel to the cell axis
throughout the movie (Fig.~\ref{fig:app_movie}a). The distribution
of filament lengths, pooled across the duration of the movie, confirms
the reported gamma distribution (Fig.~\ref{fig:app_movie}b; maximal
likelihood fits of normal, Weibull, and Rayleigh distributions yield
higher values for the Akaike information criterion \cite{Akaike1974}).
Filament length is correlated with filament weight, i.e., longer filaments
are typically thicker (Fig.~\ref{fig:app_movie}c; Pearson correlation
$p$-value $p_{\mathrm{P}}<0.05$). Moreover, the correlation between
different measures of filament curvedness, i.e., the filament bending
and the maximal filament angle, are consistently negatively correlated
with the filament length (Fig.~\ref{fig:app_movie}d; $p_{\mathrm{P}}<0.05$).

In addition to these previously analysed features of filamental organisation,
we study the course of different filament properties over time: The
average filament weight shows large fluctuations and is non-stationary
over the recording period (Fig.~\ref{fig:app_movie}e; cf.~augmented
Dickey-Fuller test $p$-value $p_{\mathrm{ADF}}\geq0.05$). This non-stationarity
suggests substantial changes in the prevalence of fine actin filament
and thick bundles, respectively, and prompts further investigations.
However, we found that the average filament length as well as the
average filament bending remain stationary over the course of $200\,\mathrm{s}$
(Fig.~\ref{fig:app_movie}e; cf.~$p_{\mathrm{ADF,length}}<0.05$
and $p_{\mathrm{ADF,conv}}<0.05$). Since the length distribution
of filaments tunes the mechanical properties of filamentous networks
\cite{Kasza2010,Bai2011}, this stationarity of the average filament
length may be of immediate biological relevance. The stationarity
of the average filament bending may be a direct consequence of the
roughly constant filament length distribution (cf.~Fig.~\ref{fig:app_movie}d)
in combination with the resultant physical constraints of actin filament
length on filament bending.

\section{\label{sec:app_overview}Supplemental Material S9: Overview of different
stages of filament decomposition of artificial, biological, and cosmic
networks}

We test our method of decomposing a given weighted network into filaments
by solving the FCP for different filamentous networks. In addition
to the four networks presented in the main text and the $100$ frames
analysed in Supplemental Material S8, we investigate four more networks
of different types and show the different stages of our analysis.
Starting from grey-scale image data of contrived, neural, cytoskeletal
and cosmic network structures (Fig.~\ref{fig:app_overview}, $1$st
column), we pre-process the images to obtain a binary representation
of the filament centre lines (Fig.~\ref{fig:app_overview}, $2$nd
column), and extract a weighted network representation as described
in the Methods (Fig.~\ref{fig:app_overview}, $3$rd column). For
the contrived and biological and the cosmic networks, we manually
assign filament identities and compute the connected components, respectively
(Fig.~\ref{fig:app_overview}, $4$th column). Finally, we decompose
the networks into filaments by solving the FCP with different options
(Fig.~\ref{fig:app_overview}, $5$th column). For the first contrived
network, we allow overlapping filaments (Fig.~\ref{fig:app_overview}a)
while for the second, grid-like contrived network (Fig.~\ref{fig:app_overview}b),
the neural networks (Fig.~\ref{fig:app_overview}c and d), the cytoskeletal
networks (Fig.~\ref{fig:app_overview}e and f), and the cosmic webs
(Fig.~\ref{fig:app_overview}g and h), we obtain exact filament covers
with options described in Fig.~\ref{fig:res_artificial}e. The agreement
of manual assignments and automated filament decompositions of the
studied networks is measured by the classical and the structure-aware
Jaccard indices $\mathrm{JI}$ and $\mathrm{JI}^{1}$ and shows good
agreement ($\mathrm{JI}^{1}$ close to $1$, and cf.~discussion of
Fig.~\ref{fig:res_actin}d) despite occasional over- (cf.~$\oplus$)
or under-segmentation (cf.~$\ominus$) of filaments. 

\begin{figure}
\centering{}\includegraphics[width=0.7\textwidth]{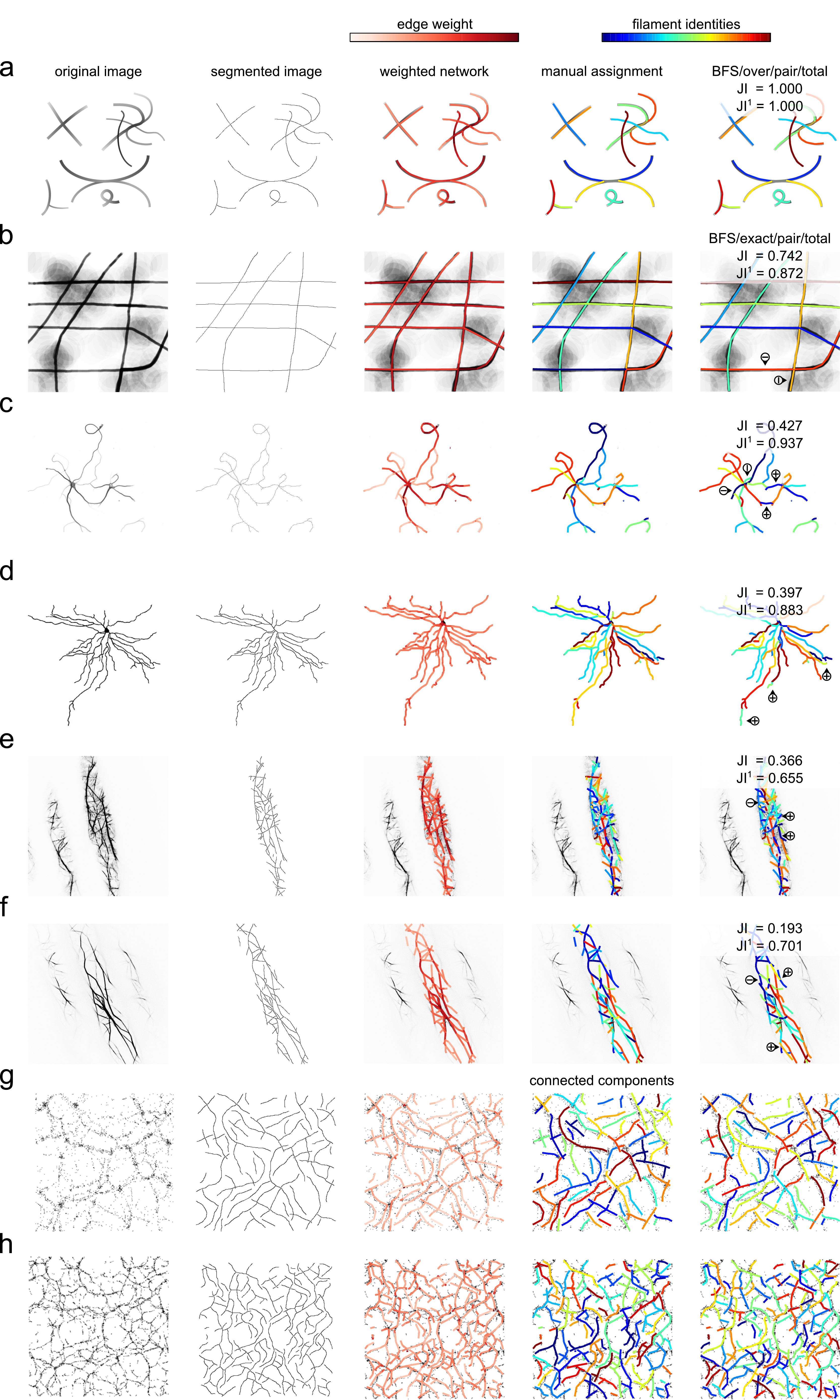}
\end{figure}

\begin{figure}
\caption{\textbf{\label{fig:app_overview}Overview of studied networks, manual
assignments, and filament covers obtained from solving the FCP.} Original
grey-scale image data ($1$st column), binary images of filament centre
lines ($2$nd column), extracted networks with colour-coded edge weights
($3$rd column), manual filament assignments of contrived and biological
networks and connected connected components of cosmic networks, respectively
($4$th column), and automatically obtained filament covers ($5$th
column). Agreement between manual decompositions and automated filament
cover is quantified by a number of measures (cf., e.g., Methods and
Fig.~\ref{fig:res_neuron}), here the classical and the structure-aware
Jaccard indices $\mathrm{JI}$ and $\mathrm{JI}^{1}$ are shown. \textbf{(a)}
Contrived network with crossing and overlapping filaments and a loop
(cf.~Fig.~\ref{fig:res_artificial}). \textbf{(b)} For a contrived,
grid-like network, the automated decomposition correctly detects most
of the filaments ($\mathrm{JI}^{1}$ close to $1$). Only the filament
in the bottom right corner with a kink is over-segmented ($\oplus$)
because the curvature restriction of the initial paths does not allow
such large angles (about $90^{\circ}$ here, cf.~Eq.~\ref{eq:supp_r_angle}).
\textbf{(c) }Neural network of hippocamal neuron (cf.~Fig.~\ref{fig:res_neuron}).
\textbf{(d)} The decomposition of the network of a retinal ganglion
cell shows good agreement with the manual results ($\mathrm{JI}^{1}$
close to $1$). A few filaments are over-segmented ($\oplus$), e.g.,
due to kinks in the filaments that are not captured by the initial
set of paths (cf.~the centre $\oplus$). \textbf{(e)} Cytoskeletal
network of actin filaments (cf.~Fig.~\ref{fig:res_actin}). \textbf{(f)}
For the actin network extracted from the confocal recording of a Lifeact-labelled
cytoskeleton, the automated partitions agrees well with the manual
results ($\mathrm{JI}^{1}$ close to $1$, and cf.~discussion of
Fig.~\ref{fig:res_actin}d). A few examples of over- and under-segmentation
($\ominus$) are marked. \textbf{(g)} Cosmic web of galaxies (cf.~Fig.~\ref{fig:res_galaxy}).
\textbf{(h)} The dense web of simulated galaxies consists of many
connected components that are further decomposed into filaments (cf.~Fig.~\ref{fig:res_galaxy}
for a discussion). Image data for panels (g) and (h) from: Stoica
et al., A\&A, 434, 423-432, 2005, reproduced with permission {\scriptsize{}\textcircled{c}}
ESO.}
\end{figure}

\section{\label{sec:app_post}Supplemental Material S10: Open contour-based
filament decomposition and filament cover-based post-processing}

Finally, we demonstrate how our filament cover-based approach may
be used to post-process and improve filament decompositions obtained
from other, e.g., open contour-based approaches. For the demonstration,
we select SOAX \cite{Xu2015}, a fully automated, stretching open
active contour-based approach which is available as an open-source
software tool to extract a network-like representation (i.e., coordinates
of filament centre lines as well as junctions are provided) from image
data. As a test case, we study the contrived filamentous structure
investigated in Fig.~\ref{fig:res_artificial}. For a fair comparison
of our and the open contour-based approach, we apply SOAX to the pre-processed
and segmented image data (cf.~Methods and Fig.~\ref{fig:app_post}a,
second panel) to which we further apply a Gaussian filter of unit
standard deviation to obtain smooth intensity gradients required by
the algorithm. SOAX is run using the default parameters and the resulting
filament identities are manually assigned to match those of the manual
solution (Fig.~\ref{fig:app_post}b). To quantify the quality of
the decomposition, we manually assign filament identities in our original
network representation (cf.~Fig.~\ref{fig:app_post}a, third panel)
according to the open contour-based result (cf.~Fig.~\ref{fig:app_post}b)
and compare the result to the manual assignment (cf.~Fig.~\ref{fig:app_post}a,
fourth panel). The structure-aware Jaccard index $\mathrm{JI}^{1}=0.938$
is close to $1$ and indicates good agreement between open-contour
based decomposition and manual filament assignment. We note that some
junctions/nodes obtained from SOAX are split in two in comparison
to our extracted networks (cf.~intersecting $\ocircle$).

Moreover severely, some filaments are over-segmented and thus fragmented
(cf.~$\oplus$), especially overlapping filaments which are not captured
in the open contour-based approach (cf.~$\ominus$). To remedy this
shortcoming, we apply our filament cover-based approach to post-process
the open contour-based decomposition and merge over-segmented filament
fragments. To this end, we convert the open contour-based filament
representation into a weighted network, where edge weights represent
average filament segment intensities as before (cf.~Methods and Fig.~\ref{fig:app_post}c).
As before, a collection of paths $\mathcal{P}^{'}$ is sampled using
a breadth-first search (\emph{BFS}) and their pairwise roughness values
$r_{p}$, $p\in\mathcal{P}^{'}$, are computed according to Eq.~\ref{eq:supp_r_pair}
(\emph{pair}). Then, to take into account the initial open contour-based
filament decomposition $\mathcal{F}$ as a starting point, in which
certain edges have already been assigned to the a given filament,
we modify the roughness values of the sampled paths: For each initial
filament or fragment that is fully contained within a sampled path,
the roughness of that path is decreased by a large value, $R_{\mathrm{\mathrm{filament}}}=10^{4}$,
which is larger than any $r_{p}$ to favour the inclusion of these
filaments or fragments in the optimal filament cover. Since the subtraction
of $R_{\mathrm{\mathrm{filament}}}$ yields negative roughness values
which would lead to the inclusion of all these paths, we add another,
even larger constant $R_{\mathrm{offset}}=10^{8}>R_{\mathrm{filament}}$
to all roughness values, i.e., 
\begin{eqnarray}
r_{p}^{'} & = & r_{p}-\sum_{\begin{array}{c}
f\in\mathcal{F}\\
f\subset p
\end{array}}R_{\mathrm{filament}}+R_{\mathrm{offset}}.\label{eq:supp_rmod}
\end{eqnarray}
For these modified roughness values $r_{p}^{'}$, we solve the FCP
by minimising the total roughness (\emph{total}) and allowing for
overlaps (\emph{over}; Fig.~\ref{fig:app_post}d). The resulting
post-processed filament decomposition merges several filament fragments
which were over-segmented by the open-contour based approach and shows
very good agreement of $\mathrm{JI}=0.776$ and $\mathrm{JI}^{1}=1.000$
with the manual filament assignment. Interestingly, in this decomposition,
parts of two filaments are interchanged (cf.~$\oplus$) as in Fig.~\ref{fig:res_artificial}f
for different FCP options. In conclusion, for any approach that detects
filaments from image data and yields a weighted network representation,
our filament cover-based approach may provide a helpful means to further
post-process and enhance the accuracy of the obtained filament decomposition.

\begin{figure}
\centering{}\includegraphics[width=0.9\textwidth]{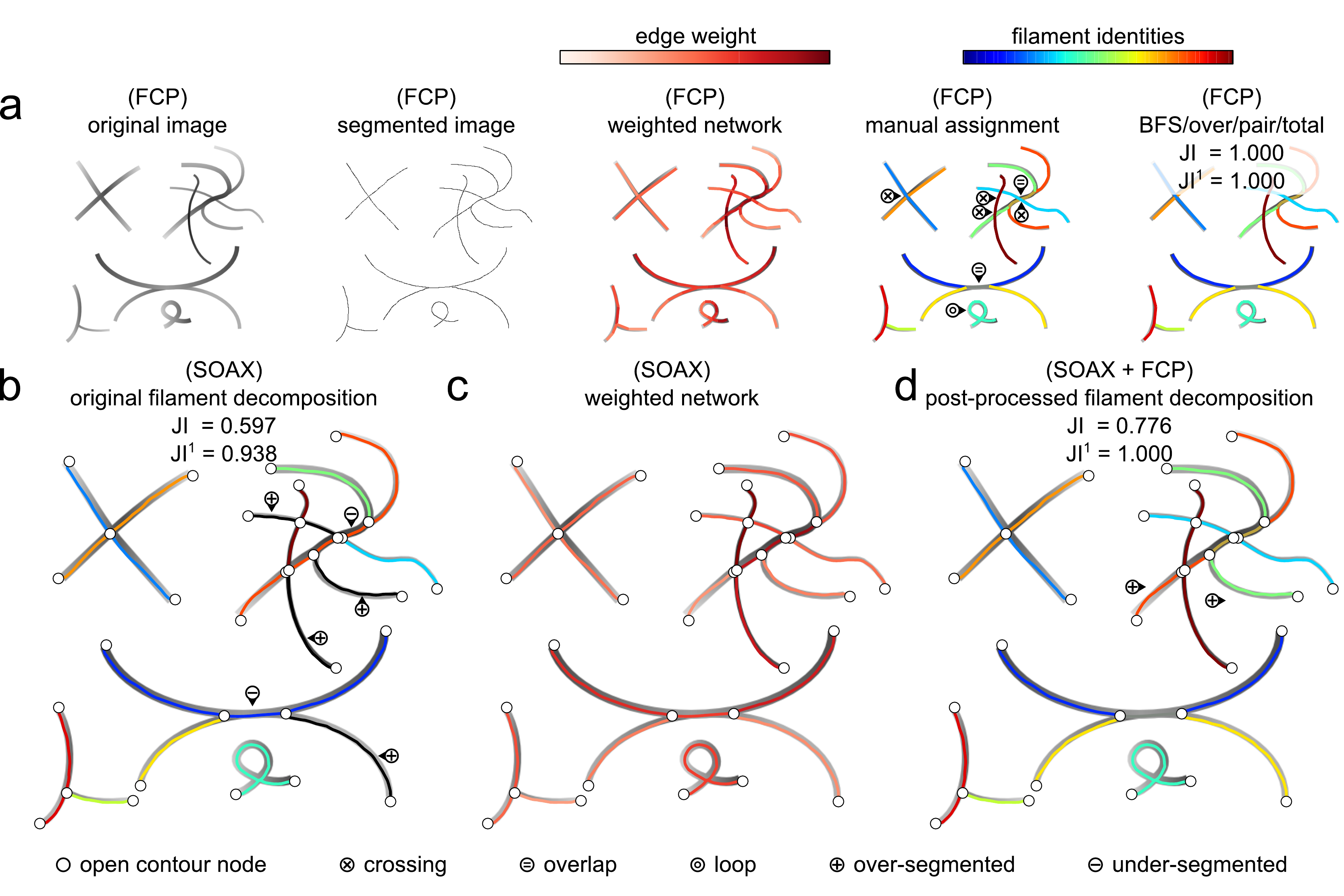}
\end{figure}

\begin{figure}
\caption{\textbf{\label{fig:app_post}Open contour-based filament detection
and filament cover-based post-processing.} \textbf{(a)} Different
stages of our filament cover problem (FCP)-based analysis for a contrived
filament structure, from original image to segmented filament centre
lines and weighted network representation, manual filament assignment
and automated solution (cf.~Fig.~\ref{fig:app_overview}a for further
explanations). \textbf{(b)} Filaments and junctions (cf.~circles)
identified from the segmented filament centre line image using SOAX,
an stretching open active contour-based approach \cite{Xu2015}. Colour-coded
filament identities were manually assigned to match those of the manual
solution in (a) and excess filament fragments were coloured black.
While the agreement with the manual solution is good ($\mathrm{JI}^{1}$
close to $1$), some filaments are over-segmented (cf.~$\oplus$)
and thus fragmented, especially at locations of filament overlaps
(cf.~$\ominus$). \textbf{(c) }Weighted network representation of
the contrived filamentous structure obtained from SOAX. \textbf{(d)
}Using the filament assignments from SOAX in (b) as a starting point,
our filament cover-based approach is used to post-process the filament
decomposition, which merges broken filament fragments and improves
the agreement with the manual solution ($\mathrm{JI}^{1}=1$).}
\end{figure}

\end{document}